\documentclass{aa} 

\usepackage{color}
\usepackage{longtable}
\usepackage{supertabular}
\usepackage{natbib}

\definecolor{linkcolor}{rgb}{0.2, 0.55, 0.9}
\usepackage[breaklinks, colorlinks, citecolor=linkcolor]{hyperref}

\usepackage[varg]{txfonts}
\usepackage{graphicx}
\usepackage{tabularx}
\usepackage{multirow}
\usepackage{booktabs}
\usepackage{natbib}
\usepackage{rotating}

\defcitealias{Umetsu+18}{U18} 

\usepackage{txfonts}

\begin{document}

\definecolor{red}{rgb}{1.,0.,0.}
\definecolor{darkgreen}{rgb}{0,0.6,0}
\definecolor{darkred}{rgb}{0.8,0,0.35}
\definecolor{grey}{rgb}{0.45,0.5,0.55}

\newcommand{\mamp}{\texttt{MAMPOSSt}}
\newcommand{\jei}{\texttt{JEI}}
\newcommand{\clumps}{\texttt{CLUMPS}}
\newcommand{\gaea}{\texttt{GAEA}}
\newcommand{\dsp}{\texttt{DS+}}
\newcommand{\rf}{\texttt{RF}}
\newcommand{\mr}{M(r)}
\newcommand{\ms}{M_{\odot}}
\newcommand{\br}{\beta(r)}
\newcommand{\bs}{\beta_{{\rm sym}}}
\newcommand{\bsz}{\beta_{{\rm sym},0}}
\newcommand{\bsi}{\beta_{{\rm sym},\infty}}
\newcommand{\bpr}{\sigma_{\rm r}/\sigma_{\theta}}
\newcommand{\rvir}{r_{200}}
\newcommand{\rtwo}{r_{-2}}
\newcommand{\cvir}{c_{200}}
\newcommand{\cnu}{c_{\nu}}
\newcommand{\vvir}{v_{200}}
\newcommand{\mvir}{M_{200}}
\newcommand{\rnu}{r_{\nu}}
\newcommand{\rb}{r_{\beta}}
\newcommand{\rs}{r_{\rm s}}
\newcommand{\ks}{\mathrm{km \, s}^{-1}}
\newcommand{\slos}{\sigma_{\rm{los}}}
\newcommand{\vrf}{v_{{\rm rf}}}
\newcommand{\fst}{f_{{\rm sub}}}
\newcommand{\fsu}{f_{{\rm sub,1}}}
\newcommand{\db}{${\rm d}\beta$}
\newcommand{\ab}[1]{\textcolor{darkred}{\bf ABi: #1}}
\newcommand{\hide}[1]{\textcolor{grey}{#1}}
\newcommand{\rev}[1]{ #1}
\newcommand{\revv}[1]{ #1}

   \title{CLASH-VLT: The variance in the velocity anisotropy profiles of galaxy clusters}

   \author{A. Biviano
          \inst{1,2}
          \and
          E. A. Maraboli
          \inst{3}
          \and
          L. Pizzuti
          \inst{4}
          \and
          P. Rosati
          \inst{5}
          \and
          A. Mercurio
          \inst{6,7}
          \and
          G. De Lucia
          \inst{1,2}
          \and
          C. Ragone-Figueroa
          \inst{8,1,2}
          \and
          C. Grillo
          \inst{3,9}
          \and
          G. L. Granato
          \inst{1,2,8}
          \and
          M. Girardi
          \inst{10,1}
          \and
          B. Sartoris
          \inst{11,1}
          \and
          M. Annunziatella
          \inst{12}}

   \institute{
    INAF-Osservatorio Astronomico di Trieste, via G. B. Tiepolo 11, 34143 Trieste, Italy\\ \email{andrea.biviano@inaf.it}
   \and
   IFPU-Institute for Fundamental Physics of the Universe, via Beirut 2, 34014 Trieste, Italy
   \and
   Dipartimento di Fisica, Università degli Studi di Milano, Via Celoria 16, I-20133 Milano, Italy
   \and
   Dipartimento di Fisica G. Occhialini, Universit\`a degli Studi di Milano-Bicocca, Piazza della Scienza 3, I-20126 Milano, Italy
   \and
   Department of Physics and Earth Sciences, University of Ferrara, Via G. Saragat, 1, 44122 Ferrara, Italy
   \and
   Dipartimento di Fisica E. R. Caianiello, Universit\`a Degli Studi di Salerno, Via Giovanni Paolo II, I-84084 Fisciano, (SA), Italy
   \and
   INAF – Osservatorio Astronomico di Capodimonte, Via Moiariello 16, I-80131 Napoli, Italy
   \and
   IATE – Instituto de Astronom\'ia Te\'orica y Experimental, Consejo Nacional de Investigaciones Científicas y T\'ecnicas de la Rep\'ublica
Argentina (CONICET), Universidad Nacional de C\'ordoba, Laprida 854, X5000BGR C\'ordoba, Argentina
   \and
   INAF -- IASF Milano, via A. Corti 12, I-20133 Milano, Italy
      \and
   Dipartimento di Fisica dell’Università degli Studi di Trieste – Sezione di Astronomia, Via Tiepolo 11, I-34143 Trieste, Italy
   \and
   University Observatory, Ludwig-Maximilians University Munich, Scheinerstrasse 1, 81679 Munich, Germany
   \and
   Centro de Astrobiología (CAB), CSIC-INTA, Ctra. de Ajalvir km 4,
Torrej\'on de Ardoz 28850, Madrid, Spain
    }

  \abstract{The velocity anisotropy profiles, $\br$, of galaxy clusters are directly related to the shape of the orbits of their member galaxies. Knowledge of $\br$ is important in order to understand the assembly process of clusters and the evolutionary processes of their galaxies, and to improve the determination of cluster masses based on cluster kinematics.}  
  {We determined the $\br$ of nine massive clusters at redshift $0.19 \leq  z \leq 0.45$ from the CLASH-VLT dataset, with $\simeq 150$ to 950 spectroscopic members each, to understand how much cluster-to-cluster variance exists in the $\br$ of different clusters and what  the main driver of this variance is.} 
  {We selected spectroscopic cluster members with the \clumps\ algorithm calibrated on cosmological simulations. We applied the \mamp\ code to the distribution of cluster members in projected phase-space to constrain the cluster mass profile, $M(r)$, using priors derived from a previous gravitational lensing analysis. Given the \mamp\ best-fit solution for $M(r)$, we then solved the inversion of the Jeans equation to determine $\br$ without assumptions of its functional form. We also ran the \dsp\ code to identify subclusters and characterize the dynamical status of our clusters.} 
  {The average $\langle \br \rangle$ is slightly radial;  the anisotropy increases from $\beta \simeq 0.2$ at the cluster center to $\beta \simeq 0.5$ at the virial radius. There is substantial variance in the $\br$ of the individual clusters that cannot be entirely accounted for by the observational uncertainties. Clusters of lower mass and with a low  concentration per given mass have more tangential $\br$ profiles. A comparison with previous works in the literature suggests that orbits are more radial in clusters at higher $z$. A comparison with cluster-sized halos in cosmological hydrodynamical simulations indicates a very good agreement for the average $\langle \br \rangle$, but a smaller variance in the profiles than observed.} 
  {Massive clusters cannot be characterized by a unique universal $\br$. The orbital distribution of cluster galaxies carries information on the merging history of the cluster.}

   \maketitle

\section{Introduction} \label{s:intro}
Determining the orbits of galaxies in clusters is useful in order to understand the evolution of clusters themselves. The initial evolution of clusters is thought to be characterized by the violent relaxation process \citep{LyndenBell67} that isotropizes galaxy orbits. Violent relaxation is also expected to occur during major mergers of clusters \citep{Valluri+07}. Since most major mergers occur with a nonzero impact parameter, transfer of the angular momentum of the secondary cluster to individual galaxies may lead to an excess of tangential orbits. In contrast, smooth accretion of galaxies from the field is instead  characterized by more radially elongated orbits \citep{LC11}. Determining the orbits of galaxies in clusters of different masses, in different dynamical states, and at different redshifts can then in principle allow us to trace the average formation and accretion history of clusters.
Moreover, galaxies are affected by different processes as they travel across a cluster. For instance, galaxies with orbits that do not come close to the cluster center are more likely to survive several pericenter passages and avoid tidal stripping by the gravitational potential of the cluster and/or gas removal by ram pressure, and they would also experience less encounters with other cluster members \citep[see, e.g.,][]{Lotz+19,Tonnesen19}.

The most common way to characterize the orbit of galaxies in clusters in observations is through the velocity anisotropy profile $\br$,
\begin{equation}
\beta = 1 - {\left\langle v_\theta^2 \right\rangle + \left\langle v_\phi^2
  \right\rangle\over 2\,\left\langle v_r^2\right\rangle} \ ,
\label{e:beta}
\end{equation}
where $\left\langle v_r^2\right\rangle$ is the mean squared radial velocity, and $\left\langle v_\theta^2 \right\rangle$ and $\left\langle v_\phi^2 \right\rangle$ are the mean squared velocity components along the two tangential directions. Observations indicate that rotational support is significant only for a minority of clusters \citep{MP17,BEN25}, so it is customary to assume no meridional streaming motions or rotation, which imply $\left\langle v_\theta^2\right\rangle = \sigma_\theta^2$ and $\left\langle v_\phi^2\right\rangle = \sigma_\phi^2$, and by symmetry, $\sigma_\phi=\sigma_\theta$. However, radial streaming motions cannot  be excluded, for instance due to infall of galaxies from the field. Radial, isotropic, and circular orbits correspond to $\beta=1, 0$, and $-\infty$, respectively. To symmetrize the range of parameter values for
radial versus tangential orbits, \citet{Mamon+19} introduced the new variable $\bs$,
\begin{equation}
    \bs \equiv \frac{\beta}{1-\beta/2},
\label{e:betasym}
\end{equation}
such that radial, isotropic, and circular orbits correspond to $\bs=2,0 $, and $-2$, respectively.

The velocity anisotropy profile $\br$ enters the Jeans equation of dynamical equilibrium \citep[e.g.,][]{BT87},
\begin{equation}
G \, M(r) = -r \, \left\langle v_r^2\right\rangle \, \left( \frac{{\rm d} \log \nu}{{\rm d} \log r} + \frac {{\rm d} \log \left\langle v_r^2\right\rangle}{{\rm d} \log r} + 2 \beta \right),
\label{e:jeans}
\end{equation}
where $\mr$ is the cluster total mass profile and $\nu(r)$ is the number density profile of the tracer of the gravitational potential. Assuming spherical symmetry, $\nu(r)$ can be derived from the observed projected number density profile, $N(R)$, of the tracer, via the Abel deprojection. However, it is impossible to infer both $\mr$ and $\br$ from the observed line-of-sight (los) velocity dispersion profile (VDP). This is called the mass-anisotropy degeneracy in the Jeans equation, which can be broken if $M(r)$ is determined with other independent probes than cluster kinematics, such as gravitational lensing and the application of the hydrostatic equilibrium to the intracluster X-ray emitting gas \citep[see, e.g.,][]{NK96,Benatov+06,HL08,Biviano+13,MBM14,Annunziatella+16,AADDV17}. 

In the absence of an external determination of $\mr$, the mass-anisotropy degeneracy can be at least partially solved by considering the entire velocity distribution of the tracers rather than restricting the analysis to the los VDP, as first pointed out by \citet{Merritt87}. \citet{LM03} used the kurtosis profile of the velocity distribution of Coma cluster galaxies, in addition to the los VDP, to infer mostly isotropic orbits for its member galaxies. Their method was validated by \citet{Sanchis+04}, using cosmological N-body simulations. \citet{Wojtak+08,Wojtak+09} constructed distribution function models for the energy and angular momentum of the system to constrain the orbital distribution of cluster galaxies by using their full projected phase-space distribution. \citet{WL10} applied this method to nearby clusters and found an average $\br$ close to isotropic near the center and mildly more radial outside, but with substantial variance from cluster to cluster. 

Another method that uses the full projected phase-space distribution of cluster galaxies to infer $\br$ and at the same time $\mr$ is \mamp\footnote{The code of an extended version of \mamp, that also includes alternative models to general relativity, \texttt{MG-MAMPOSSt} \citep{Pizzuti+23}, is available at \texttt{https://github.com/Pizzuti92/MG-MAMPOSSt}.} \citep{MBB13}. In \mamp\ the distribution function is no longer expressed in terms of energy and angular momentum, but in terms of the three-dimensional velocity distribution function, assumed to be Gaussian. Using parametric forms for $\mr$ and $\br$, \mamp\ solves the Jeans Eq.~(\ref{e:jeans}) to compute the likelihood of the distribution of tracers in projected phase-space. \mamp\ has been validated with N-body cosmological simulations  \citep{MBB13,Tagliaferro+21,Read+21}.

Similarly to what was found by the distribution function method of \citet{WL10}, the \mamp-based studies found nearly isotropic orbits near the cluster center and increasingly radial orbits at larger cluster-centric distances. Red and blue galaxies share a similar orbital distribution in $z \gtrsim 0.2$ clusters \citep{Biviano+13,Mercurio+21,Biviano+16,Biviano+17a,Capasso+19}, while
in low-$z$ clusters the orbits of early-type galaxies are less radial than those of late-type galaxies \citep{Mamon+19,VR25}.
Extremely elongated radial orbits characterize galaxies subjected to ram-pressure stripping \citep{Biviano+24}. Galaxies in fossil systems appear to be characterized by more radially elongated orbits than their counterparts in normal clusters \citep{Zarattini+21}.

Another way to at least partially solve the mass anisotropy degeneracy is to use several independent tracers of the same potential \citep{Battaglia+08}.  \citet{BK04} adopted the $\mr$ derived by \citet{KBM04}, who used only early-type galaxies as tracers,
to invert the Jeans equation by the method of \citet[][first developed by \citeauthor{BM82} \citeyear{BM82}]{SSS90}, 
and determined $\br$ for the cluster galaxy populations that were not used in the determination of $\mr$. By a similar approach, \cite{Adami+09} determined $\beta$ for dwarf galaxies in Coma,
adopting the $\mr$ derived by \citet{GDK99}, who used bright galaxies as tracers. \citet{BP09} solved the Jeans equation separately for passively evolving galaxies and emission line galaxies in two stacks of clusters at mean redshifts of 0.07 and 0.56, and found suggestive evidence of the orbital evolution of the passive galaxies from radial to isotropic with decreasing $z$.

Due to the lack of sufficient statistics for each individual cluster, to date most cluster $\br$ determinations have  been based on stacked samples \citep{BK03,BP09,Biviano+16,Biviano+17a,Capasso+19,Mamon+19}. Individual cluster studies show a variety of $\br$
\citep{NK96,LM03,Benatov+06,HL08,Biviano+13,MBM14,Annunziatella+16,AADDV17,Mercurio+21}, but it is difficult to know how much of this variance is due to the different methodological approaches used by the different authors. Nevertheless, significant $\br$ variance is also found in two studies that have determined $\br$ for several clusters in homogeneous ways, those of \citet[][41 $z<0.1$ clusters with $\geq 66$ members each]{WL10} and \citet[][ten $z<0.1$ clusters with $\geq 75$ members each]{Li+23}.

The aim of this paper is to better constrain the cluster-to-cluster variance in $\br$, using a sample of nine very massive clusters ($\mvir$\footnote{$M_\Delta$ is the mass contained in a sphere of radius $r_\Delta$ with a mean density equal to $\Delta$ times the critical density of the Universe at the cluster redshift. We also define $v_{\Delta} \equiv (G M_{\Delta}/r_{\Delta})^{1/2}$, where $G$ is the gravitational constant.} $> 0.7 \times 10^{15} \, \ms$) located at higher $z$ ($0.19-0.45$) than the clusters analyzed by \citet{WL10} and \citet{Li+23}. Our nine clusters are part of the CLASH-VLT sample (\citeauthor{Rosati+14} \citeyear{Rosati+14}, Rosati et al. in prep.), each with $\simeq 150$ to 950 spectroscopic members in the virial region, thus allowing the determination of $\br$ with much better statistics than the previous studies. In our analysis we first
determine the nine cluster $\mr$ profiles with \mamp, using priors obtained from gravitational lensing \citep[][U18 hereafter]{Umetsu+18}, and then invert the Jeans equation by the technique of \citet{SSS90} and \citet{DM92} to determine the individual cluster $\br$ profiles and a weighted mean of these profiles. We find significant variance in the individual $\br$ profiles, and try to understand its origin by correlating $\br$ deviations from the mean with other cluster properties. We compare our results with the $\br$ of halos in two cosmological simulations, and with the results of \citet{WL10} and \citet{Li+23} for nearby clusters.

The structure of this paper is as follows. In Sect.~\ref{s:data} we describe the dataset, and in Sect.~\ref{s:anal} the method of analysis. In Sect.~\ref{s:resu} we present our results, the $\br$ of the nine clusters, and their weighted average. We then explore the correlations between the $\br$ deviations from the average and other cluster properties, and compare our results to those of \citet{WL10} and \citet{Li+23} at lower $z$, and to the $\br$ of halos from cosmological simulations. In Sect.~\ref{s:disc} we discuss our results and in Sect.~\ref{s:conc} we provide our summary and conclusions.
Throughout this paper we assume a $\Lambda$CDM cosmology with $\Omega_{\rm m,0}=0.3$, $\Omega_{\Lambda,0}=0.7$, and $H_0 = 70 \,\rm km \,s^{-1} \, Mpc^{-1}$.
\begin{table*}
\centering
\caption{Cluster sample.}
\label{t:data}
\begin{tabular}{llrrccc|cc} 
\toprule
\multirow{2}{*}{Cluster name}
& \multirow{2}{*}{Short name} & \multirow{2}{*}{RA} & \multirow{2}{*}{Dec} & \multirow{2}{*}{$\overline{z}$} & \multirow{2}{*}{m$_{\rm{R,lim}}$} & \multirow{2}{*}{N$_{{\rm m}}$} & \multicolumn{2}{c}{U18} \\
& & & &  & & & $\rvir$ & $\rs$ \\
\midrule
Abell 209                    & A209   &  22.9869 & $-$13.6112 & 0.209 & 23.0 & 954 & $2.40 \pm 0.15$ & $0.70 \pm 0.15$ \\
Abell 383                    & A383   &  42.0142 &  $-$3.5291 & 0.187 & 23.5 & 485 & $1.95 \pm 0.27$ & $0.78 \pm 0.51$ \\
MACS0329.7$-$0211              & M329   &  52.4232 &  $-$2.1961 & 0.450 & 23.0 & 262 & $1.90 \pm 0.11$ & $0.35 \pm 0.09$ \\
MACS1115.9+0129              & M1115  & 168.9662 &   1.4986 & 0.352 & 23.0 & 472 & $2.22 \pm 0.16$ & $0.89 \pm 0.26$ \\
MACS1206.2$-$0847              & M1206  & 181.5506 &  $-$8.8009 & 0.440 & 23.0 & 409 & $2.02 \pm 0.14$ & $0.35 \pm 0.10$ \\
MACS1931.8$-$2635              & M1931  & 292.9567 & $-$26.5758 & 0.352 & 22.5 & 250 & $1.92 \pm 0.16$ & $0.25 \pm 0.06$ \\
MS2137$-$2353                  & MS2137 & 325.0632 & $-$23.6612 & 0.313 & 23.0 & 159 & $1.90 \pm 0.19$ & $0.79 \pm 0.34$ \\
RXJ2129.7+0005               & R2129  & 322.4165 &   0.0892 & 0.234 & 23.0 & 248 & $1.75 \pm 0.18$ & $0.60 \pm 0.26$ \\
RXJ2248.7$-$4432 (Abell S1063) & R2248  & 342.1832 & $-$44.5309 & 0.348 & 23.5 & 905 & $2.30 \pm 0.23$ & $1.44 \pm 0.64$ \\
\bottomrule
\end{tabular}
\tablefoot{RA and Dec are the right ascension and declination of the BCG in degrees. N$_{{\rm m}}$ is the number of members with magnitude m$_\mathrm{R} \leq$ m$_{\mathrm{R,lim}}$
in the radial range $0.05 \, \mathrm{Mpc} \leq R \leq 1.36 \, \rvir$. The U18 values of $\rvir$ and $\rs$ are derived from $\mvir$ and $\cvir$ as listed in Table~1 of \citetalias{Umetsu+18}.  Error bars are 1$\sigma$.
Radii are given in units of Mpc. }
\end{table*}

\section{The dataset}\label{s:data}
Our dataset consists of nine massive galaxy clusters ($\mvir > 0.7 \times 10^{15} \, \ms$)  at intermediate redshifts ($0.19 \lesssim z \leq 0.44$) from the Cluster Lensing And Supernova survey with Hubble \citep[CLASH; see][]{Postman+12}, with extensive spectroscopic follow-up at the ESO VLT (the CLASH-VLT sample; \citeauthor{Rosati+14} \citeyear{Rosati+14}, Rosati et al. in prep.). In Table \ref{t:data} we list these nine cluster identification names, the short names we use in this paper, their center coordinates, and mean $z$. These nine clusters are a subset of the original CLASH-VLT sample of 13. The excluded clusters are MACS1311.0$-$0310 because it does not have
a $\mr$ determination by the gravitational lensing analysis of \citetalias{Umetsu+18}; MACS0416.1$-$2403 because it is a merging cluster with a very complex dynamical state \citep{Balestra+16}, which probably invalidates the application of the Jeans equation; RXJ1347.5$-$1145 because it lacks complete photometric coverage to fully assess its spectroscopic completeness; and MACS2129.4$-$0741 because of insufficient spectroscopic data.

In Table~\ref{t:data} we also list the number of spectroscopic members in the virial region used in the dynamical analysis of this paper (N$_{{\rm m}}$, see below for the membership definition), and brighter than the R-band magnitude limits, m$_{\mathrm{R,lim}}$, listed in Table \ref{t:data}. These magnitude limits were set to ensure a spectroscopic completeness $\geq 30$\% throughout the whole virial region. The completeness estimates are described in Maraboli et al. (in prep.). They are based on color-color cuts in the m$_\mathrm{B}$-m$_\mathrm{R}$ versus m$_\mathrm{V}$-m$_\mathrm{I}$ plane that allow  both red and blue cluster galaxies to be included, and on color-magnitude cuts in the B$-$R versus R  plane that allow  low-$z$ bright nonmembers to be rejected. Previous works exploring the dynamics of CLASH-VLT clusters \citep{Annunziatella+16,Sartoris+20,Mercurio+21,Biviano+23,Girardi+24}
quote larger numbers of spectroscopic members since they did not apply these color and magnitude cuts to the sample.
Finally, in Table~\ref{t:data} we list the \citetalias{Umetsu+18} estimates of their virial and scale radius, $\rvir$ and $\rs$, that we computed from the values of $\mvir$ and $\cvir$\footnote{The concentration $\cvir \equiv \rvir/\rs$, where $\rs$ is the scale parameter of the NFW profile \citep{NFW96,NFW97}.} listed in Table~1 of \citetalias{Umetsu+18}.

Photometric data were obtained with the Suprime-Cam imager \citep{Miyazaki+02} at the Subaru telescope, for all nine clusters except for R2248, which was observed with the Wide-Field Imager at the ESO 2.2 m MPG / ESO telescope \citep{Gruen+13,Mercurio+21}. We refer to \citet[][Sect.~4.2 and Table 1 and 2]{Umetsu+14} for a detailed description of the available multiband images for each of the nine clusters. 

The spectroscopic data come mostly from the CLASH-VLT program\footnote{VLT program identification numbers 60.A-9345, 095.A-0653(A), 097.A-0269(A), 186.A-0798} (\citeauthor{Rosati+14} \citeyear{Rosati+14}, Rosati et al. in prep.), and were obtained with the VIsual Multi-Object Spectrograph \citep[VIMOS,][]{LeFevre+03}. Part of these spectroscopic data have already been released (see \citealt{Annunziatella+16} for A209, \citealt{Biviano+13,Caminha+17} for M1206, and \citealt{Karman+17,Mercurio+21} for R2248). Additional spectroscopy, deeper but in smaller fields, has been obtained for some of our clusters with the Multi Unit Spectroscopic Explorer \citep[MUSE,][]{Bacon+10}. The full spectroscopic dataset contains 24100 high-quality $z$ values. Further details on the photometric and spectroscopic datasets are given in Maraboli et al. (in prep.).

\section{Analysis}\label{s:anal}
In our analysis we adopt the spherical approximation. Although clusters are not spherically symmetric, this is a necessary approximation if we want to solve the spherical Jeans equation. Going beyond the spherical approximation requires introducing other parameters that are impossible to tightly constrain even with the unprecedented amount of spectroscopic data available for our dataset. As we discuss in Sect.~\ref{ss:simu}, we adopt the same spherical approximation to derive the $\br$ of cluster-size halos from numerical simulations, to allow a proper comparison with our results. However, the spherical approximation was not adopted in the gravitational lensing analysis of \citetalias{Umetsu+18}, who modeled the cluster mass profiles with elliptical NFW \citep{NFW96,NFW97} models. Since we used the values derived from \citetalias{Umetsu+18} for $\mvir$ and $\cvir$ as priors in our dynamical analysis with \mamp, we compared the elliptical 3D mass profiles of \citetalias{Umetsu+18} with the 3D mass profiles derived by the gravitational lensing analysis of \citet{Umetsu+14} under the spherical approximation. Specifically, we computed the ratios of these elliptical and spherical
mass profiles, $M_e/M_s$,
at three different overdensities, $\Delta=500, 200,$ and 100. We found an average $M_e/M_s = 0.97$ over the nine clusters in our sample, the same value at all $\Delta$ values. The rms of these ratios are 0.08, 0.13, and 0.17 at $\Delta=500, 200,$ and 100, respectively. For none of our nine clusters does the $M_e/M_s$ ratio differ  from unity at more than 1.6$\sigma$ at any $\Delta$. We conclude from this analysis that
the spherical approximation does not cause significant differences in the lensing $\mr$ determinations. 

Since our analysis is based on the Jeans equation, we also implicitly assume dynamical equilibrium, a valid assumption if the number of galaxies in any given region of the cluster phase-space does not change. Rapid mass growth of a cluster can invalidate this assumption. Based on the predictions of the theoretical model by \citet{Li+07}, it is expected that a cluster with the mean mass and the mean redshift of our sample has grown in mass by $\sim 20$\% over the past $\sim 1$~Gyr, which is the dynamical time of the cluster \citep{Sarazin86}. This is not entirely negligible, and given the substantial variance in the mass accretion history of different halos of similar mass, we cannot exclude that some of our clusters depart from dynamical equilibrium due to a recent major merger. We come back to this question in Sect.~\ref{ss:corr} when we consider the possible effect of rich subclusters on the $\br$.

Since accretion is most likely to occur inside-out, in order to minimize its effect on the cluster dynamical state we restricted our dynamical analysis to the $R \leq 1.36 \, \rvir$ region. At the mean $z=0.32$ of our clusters, $1.36 \, \rvir$ is  between the virial radius $\sim r_{126} \approx 1.22 \, \rvir$ \citep{BN98}, and the splashback radius $\sim 1.5 \, \rvir$, within which the cluster dynamics is still dominated by galaxies orbiting the cluster potential \citep{PGKD24}.
We also excluded the central 50 kpc region from our analysis since this region is dominated by the gravitational potential of the brightest cluster galaxy, so the total $\mr$ deviates from the NFW model \citep[see, e.g.,][]{BS06,Mamon+19,Sartoris+20}.

\subsection{Cluster membership}
\label{ss:members}
We used the CLUster Membership in Phase Space (\clumps) method of \citet{Biviano+21} to establish cluster membership, with parameters calibrated on light cones constructed from \gaea\ mocks \citep{DeLucia+24}. The GAlaxy Evolution and Assembly (\gaea) semi-analytical model is a substantial update of the model of \citet{DLB07}, and is based on the Millennium simulation \citep{Springel+05}. We considered the 358 richest halos with more than 150 members to mimic the richness distribution of our nine clusters. We calibrated the \clumps\ parameters, d$R$, and d$V$ to optimize the membership completeness and purity of simulated galaxies in halos. We found that 
d$R=450$ kpc and d$V=200 \, \ks$ give an average membership completeness $>0.99$ and a purity $>0.85$. The values of the \clumps\ parameters we find differ from those recommended by \citet{Biviano+21}, but in that study the mock sample contained halos with as few as ten members. 

\begin{figure}
\centering
\includegraphics[width=\hsize]{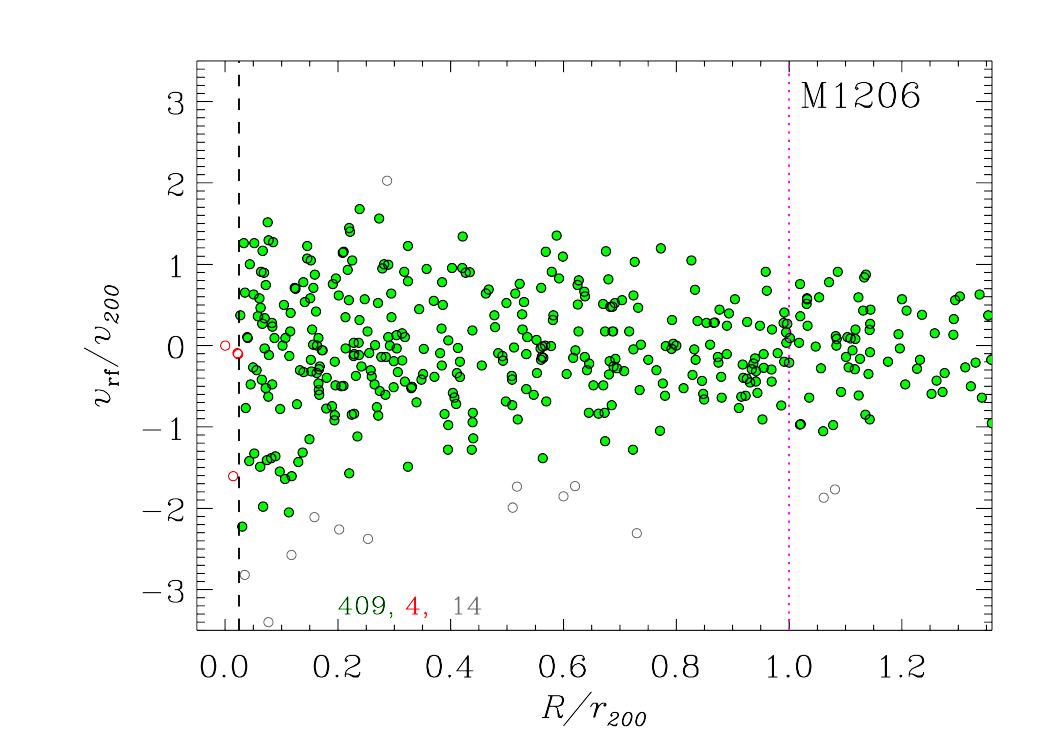}
\caption{Projected phase-space diagram of the 427 galaxies selected as members of the M1206 cluster (other clusters are shown in Fig.~\ref{f:pps}) by the \clumps\ method in the region $R \leq 1.36 \,\rvir$. The three numbers at the bottom left indicate, from left to right, the number of galaxies selected for the dynamical analysis
(black-circled green dots); the number of \clumps\ members that are excluded from the dynamical analysis because they are at $R \leq 0.05$ Mpc, which is the region dynamically dominated by the BCG (red circles), and the number of \clumps\ members that are not used in the dynamical analysis because they are flagged as interlopers by the escape-velocity criterion (gray circles). The vertical lines indicate $R=0.05$ Mpc and $R=\rvir$.  The values of $\rvir$ and $\vvir$ come from \citetalias{Umetsu+18}.} 
\label{f:pps1}
\end{figure}

We ran \clumps\ on our nine clusters, using the positions of their BCG as a center (the BCG coordinates are given in Table~\ref{t:data}). 
After running \clumps\ we further refined the membership selection by taking advantage of the fact that we already had good estimates of the cluster $M(r)$ from the gravitational lensing analysis of \citetalias{Umetsu+18}. Given $M(R)$, we computed the escape velocity $v_{\rm esc}(R)$ as a function of projected radius, $R$, by considering a wide range of possible $\br$. Specifically, we considered  two models:
\begin{equation}
    \beta(r)=\beta_0 + (\beta_\infty-\beta_0)\frac{r^\delta}{r^\delta+r_\beta^\delta},
    \label{e:betamodels}
\end{equation}
with $\delta=1$ \citep[generalized Tiret model, gT hereafter][]{Tiret+07} and $\delta=2$ \citep[generalized Osipkov-Merritt model, gOM hereafter][]{Osipkov79,Merritt85}, where we forced $r_\beta \equiv r_s$, based on the results of numerical simulations \citep{MBM10}, to limit the number of free parameters. For each of the two models, we considered three values for $\beta_0$, $-$99, 0, and 0.99 (which correspond to $\bs$ values of $-$1.96, 0, and 1.96, respectively) and the same three values for $\beta_\infty$, for a total of 18 profiles. For each of these $\br$, we calculated $v_{\rm esc}(R)$ and excluded from the sample of cluster members any galaxy with a los velocity in the cluster rest-frame, $\lvert \vrf \rvert > {\rm max}(v_{\rm esc}(R))$, where ${\rm max}(v_{\rm esc}(R))$ is the largest escape velocity at the galaxy projected radius, $R$, among the 18 considered. This procedure ensured that we removed any remaining interlopers after the \clumps\ selection in a conservative way, without biasing our results, given the extremely wide range of $\br$ considered. On average, this additional membership selection removed only 4\% of the galaxies initially identified as members by \clumps.

In Fig.~\ref{f:pps1} we show the distribution in projected phase-space of member galaxies in the cluster M1206, selected by the \clumps\ algorithm and by the $v_{\rm esc}(R)$ condition (all nine clusters are shown in Fig.~\ref{f:pps}). In Table~\ref{t:data} we list the number of members N$_{{\rm m}}$ selected by \clumps\ and by the $v_{\rm esc}(R)$ condition,
in the radial range we consider in our dynamical analysis, i.e.,
from 50 kpc to $1.36 \, \rvir$, where $\rvir$ comes from \citetalias{Umetsu+18}. 

Previous dynamical analysis were conducted on four of the nine clusters in our sample, based on the same dataset, but using different membership selection techniques, and no magnitude cut \citep{Biviano+13,Annunziatella+16,Sartoris+20,Mercurio+21,Girardi+24}. In Fig.~\ref{f:vdpcfr} we show that the different membership selections do not significantly modify the los velocity dispersion profiles of these four clusters. This suggests that our results are stable versus different membership selections.
\begin{table*}
\centering
\caption{Results from the $N(R)$ fits and \mamp.}
\label{t:mamp}
\setlength{\extrarowheight}{0.2cm}
\begin{tabular}{llccclccr} 
\toprule
Short name & $N(R)$ & $\rnu$ & $\rvir$ & $\rs$ & $\br$ & $\bs(0.05)$ & $\bs(r_{200})$ & $\chi^2$\\
\midrule
A209   & King & $0.66_{-0.02}^{+0.02}$ &   $2.31_{-0.05}^{+0.05}$ & $0.66_{-0.09}^{+0.10}$ & gOM & $[-1.1, -0.4]$ & $[ 0.8,  1.4]$ & 13.7 \\
A383   & NFW  & $0.46_{-0.06}^{+0.06}$ &   $1.83_{-0.06}^{+0.06}$ & $0.24_{-0.11}^{+0.13}$ & gOM & $[-1.2,  1.3]$ & $[ 0.3,  1.3]$ &  4.3 \\
M329   & NFW  & $0.70_{-0.12}^{+0.15}$ &   $1.84_{-0.07}^{+0.07}$ & $0.33_{-0.09}^{+0.07}$ & gOM & $[-1.4,  1.1]$ & $[-0.5,  1.0]$ &  2.0 \\
M1115  & NFW  & $1.06_{-0.13}^{+0.18}$ &   $1.69_{-0.17}^{+0.14}$ & $0.93_{-0.36}^{+0.49}$ & gOM & $[-1.0,  1.6]$ & $[-0.5,  0.4]$ &  1.4 \\
M1206  & NFW  & $0.67_{-0.10}^{+0.12}$ &   $2.06_{-0.06}^{+0.07}$ & $0.25_{-0.09}^{+0.11}$ & gOM & $[-0.8,  1.5]$ & $[ 0.2,  1.0]$ &  4.4 \\
M1931  & NFW  & $1.13_{-0.23}^{+0.28}$ &   $1.91_{-0.07}^{+0.07}$ & $0.24_{-0.06}^{+0.06}$ & gT  & $[-1.2,  0.6]$ & $[-0.1,  1.4]$ &  4.0 \\
MS2137 & NFW  & $1.09_{-0.25}^{+0.32}$ &   $1.70_{-0.10}^{+0.11}$ & $0.46_{-0.23}^{+0.29}$ & gOM & $[-1.3,  1.6]$ & $[-0.8,  1.1]$ &  9.8 \\
R2129  & NFW  & $1.02_{-0.21}^{+0.26}$ &   $1.75_{-0.09}^{+0.08}$ & $0.46_{-0.22}^{+0.28}$ & gOM & $[-1.3,  1.5]$ & $[-0.9,  0.3]$ & 12.4 \\
R2248  & NFW  & $0.97_{-0.09}^{+0.10}$ &   $2.40_{-0.08}^{+0.08}$ & $0.76_{-0.21}^{+0.26}$ & gOM & $[-0.9,  0.2]$ & $[ 0.6,  1.4]$ &  5.1 \\
\bottomrule
\end{tabular}
\tablefoot{Radii are given in units of Mpc. 
All error bars are marginalized 1$\sigma$. We list the lower and upper 1$\sigma$ limit on $\bs(0.05)$ and $\bs(r_{200})$. 
The values of $\chi^2$ are derived from the fit of the \mamp\ solution to the observed los VDP computed in seven radial bins.}
\end{table*}

\subsection{Number density profiles}\label{ss:NR}

The 3D number density profile of cluster members, $\nu(r)$, enters the Jeans Eq.~\ref{e:jeans}. Under the spherical assumption, this can be directly obtained from the projected number density profile, $N(R)$, via the Abel inversion \citep[see, e.g.,][]{BT87}. We determine $N(R)$ using the radial distribution of cluster members after correcting for spectroscopic incompleteness, as described in Maraboli et al. (in prep.). We determine $N(R)$ in N$_{\rm m}^{1/2}$ bins, where N$_{\rm m}$ is the number of cluster members in the radial range $0.05 \, \rm{Mpc} \leq R \leq 1.36 \, \rvir$, with $\rvir$ from \citetalias{Umetsu+18}. This binned profile is used in the Jeans Equation Inversion (\jei) analysis described in Sect.~\ref{ss:jeans}. We also provide model fits to the $N(R)$ using a maximum likelihood procedure that does not require binning. We consider two models, NFW \citep[in projection, see][]{Bartelmann96} and \citet{King62_denslaw}, each characterized by just one free parameter, the scale radius $\rnu$. The normalization of $N(R)$ is not a free parameter in the fit since it is set by the requirement that the integrated profile corresponds to the number of cluster members corrected for incompleteness. 

We list in Table~\ref{t:mamp} the best-fit models and scale radii for our nine clusters. Only in one case (A209) is the King model, characterized by a central core, a better fit to $N(R)$ than the NFW model. In all cases the models provide acceptable fits to the observed profiles, within the 90\% confidence level, according to a $\chi^2$ test.

\subsection{Solving the Jeans Equation for $\beta(r)$}\label{ss:jeans}

To determine $\br$ from the observables, we first need to break the Jeans equation degeneracy using a $\mr$ estimate. For this we use the
\mamp~ method of \citet{MBB13} applied to the phase-space distribution of cluster members, using the public code \texttt{MG-MAMPOSSt} \citep{Pizzuti+23}. \mamp~ estimates the probability of finding a cluster galaxy at its observed position in projected phase-space, for a set of $\mr$ and $\br$ model parameters, by solving the spherical Jeans equation, under the assumption of a Gaussian 3D velocity distribution. The best-fit $\mr$ and $\br$ parameters are found by maximizing the product of the individual galaxy probabilities. To improve the uncertainties in the posteriors of the \mamp\ analysis we adopted the $\mr$ estimate by \citetalias{Umetsu+18} as priors whenever possible, that is, for eight of our nine clusters, as described below.

In a first iteration we ran \mamp\ by adopting the NFW model for $\mr$, using flat priors on the $\rvir$ and $\rs$ parameters in the ranges 1.0--3.0 and 0.2-2.0 Mpc, respectively. We used a Gaussian prior for the scale parameter of the $\nu(r)$ profile, $\rnu$, with a mean and 1$\sigma$ error obtained from the maximum likelihood fit to the completeness-corrected $N(R)$ (see Sect.~\ref{ss:NR} and Table~\ref{t:data}). We considered two $\br$ models, gT and gOM (see eq.~\ref{e:betamodels}) with flat priors on their $\beta_0, \beta_{\infty}$ parameters in the range $-$3 to 0.96 (which corresponds to $\bs$ values of $-$1.2 and 1.85, respectively). We then ran a  Markov chain Monte Carlo (MCMC) procedure with 25,000 steps per cluster, using the \citet{GR92} criterion to check for convergence, adopting a threshold of $\hat{R}=1.01$ for the Gelman-Rubin coefficient.

We find that the posteriors for $\rvir$ and $\rs$ obtained with \mamp\ are in agreement at better than 1.5$\sigma$ with the values given by \citetalias{Umetsu+18} (and listed in Table~\ref{t:data}) for eight of our nine clusters. For M1115 we find a best-fit value of $\rvir=1.69_{-0.17}^{+0.14}$ Mpc, different from the \citetalias{Umetsu+18} value at 2.5$\sigma$ and in better agreement with the results of the X-ray analysis of \citet{Donahue+14}. For the remaining eight clusters we ran  the \mamp\ analysis a second time, this time using Gaussian priors on the $\rvir$ and $\rs$ parameters with mean and 1$\sigma$ uncertainties as given by \citetalias{Umetsu+18} (and listed in Table~\ref{t:data}). 

The resulting best-fit parameters and their 1$\sigma$ marginalized errors are listed in Table~\ref{t:mamp}. For the $\br$ profiles we list the values of $\bs$ at $r=0.05$ Mpc and $r_{200}$ since they are more directly comparable to previous results than the $\beta_0, \beta_{\infty}$ parameters. 
The $\br$ profiles of the clusters A209, M1206, and R2248 are consistent with the previous \mamp\ determinations by \citet{Annunziatella+16}, \citet{Biviano+13}, and \citet{Mercurio+21}, respectively.

\begin{figure}
\centering
\includegraphics[width=\hsize]{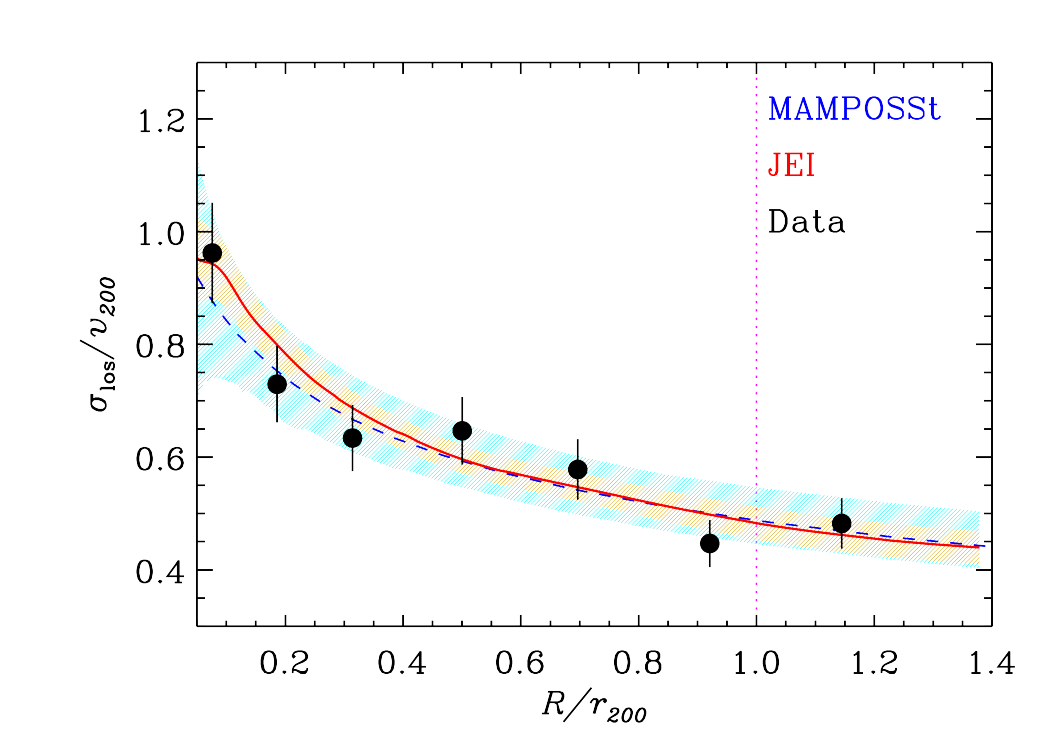}
\caption{Dots with 1$\sigma$ error bars: los VDP of M1206. Dashed blue line and cyan shading: Best-fit \mamp\ solution within 68\% confidence levels, estimated on a random selection of 3000 MCMC steps. Red solid line and orange shading: \jei-predicted VDP and 68\% confidence levels. The values of $\rvir$ (indicated by the vertical magenta line) and $\vvir$ are the best fits of the \mamp\ analysis (other clusters are shown in Fig.~\ref{f:vdp}.)}
\label{f:vdp1}
\end{figure}

 The best fit is not necessarily an acceptable fit to the data; to
check \mamp~ best-fit solutions, we compared them to the observables, the los VDP in this case, evaluated in seven radial bins. The $\chi^2$ values of this comparison are given in Table~\ref{t:mamp}. They are computed as 
\begin{equation}
\chi^2 = \Sigma_{i=1}^7 (\Delta \sigma_{{\rm los},i})^2/\delta^2_{\sigma,i},
\end{equation}
where $\Delta \sigma_{{\rm los},i}$ is the difference between the observed and predicted los velocity dispersion at the radial bin $i$, and $\delta_{\sigma,i}$ is the observational uncertainty on the observed velocity dispersion. The comparison between predicted and observed VDPs is shown in Fig.~\ref{f:vdp1} for the cluster M1206 as an example; all  nine clusters are shown in Fig.~\ref{f:vdp}.

\mamp\ results were not obtained by fitting directly the VDP, so we do not expect perfect agreement, and yet the $\chi^2$ values of two clusters (A209 and R2129) are rather high. This could either indicate violation of the dynamical equilibrium condition or that the chosen $\br$ models are too rigid to represent the real $\br$. While there is a large consensus on the fact that the NFW model is a good representation of $\mr$ on the cluster scale, at least in the radial range considered in this work, no claim for the universality of $\br$ has ever been made. For this reason, we proceeded with the second part of our analysis to determine $\br$ free of model assumptions, using the \jei~ method. 

The \jei\ method solves the Jeans equation for $\br$, given $\mr$. It was first developed by \citet{BM82}; here we followed the implementation of the method by \citet{SSS90} and \citet{DM92}. The method we briefly outline here has already been applied to several datasets with small variations \citep[e.g.,][]{BK04,Biviano+13,Annunziatella+16,Biviano+16,Zarattini+21,Biviano+24}. 

In our \jei\ analysis, we adopted the NFW $\mr$ with the \mamp\ best-fit parameters (Table~\ref{t:mamp}). The observables in the \jei\ method are the binned $N(R)$ (see Sect.~\ref{ss:NR}) and the binned VDP (described above). We smoothed both profiles using the \texttt{LOWESS} smoothing technique \citep[see, e.g.,][]{CMG84,Gebhardt+94} with a smoothing length of 0.7. We extrapolated the smoothed profiles to large radii (30 Mpc) following \citet{Biviano+13} to allow  the equations that contain integrals up to infinity to be solved. The resulting $\br$ is a nonparametric profile. 

The uncertainties on $\br$ are estimated as follows. We run the \jei\ procedure 1500 times. In each run, we 
\begin{itemize}
\item bootstrap resample the projected phase-space (cluster-centric distances and  rest-frame velocities), and determine new $N(R)$ and VDP on the bootstrap sample;  
\item randomly select the \mr\ parameters from one of the MCMC steps of the \mamp\ analysis;
\item choose a random value for the \texttt{LOWESS} smoothing parameter between 0.5 and 0.9.
\end{itemize}
At the end we define the intervals containing 68\% of the $\br$ profiles at 50 points uniformly spaced in the radial interval analyzed 0.05--1.36 $\rvir$. We check for convergence by comparing the 68\% confidence region we derive, with that obtained with 1000 runs.

\section{Results}\label{s:resu}
\begin{figure*}
\centering
\includegraphics[width=\hsize]{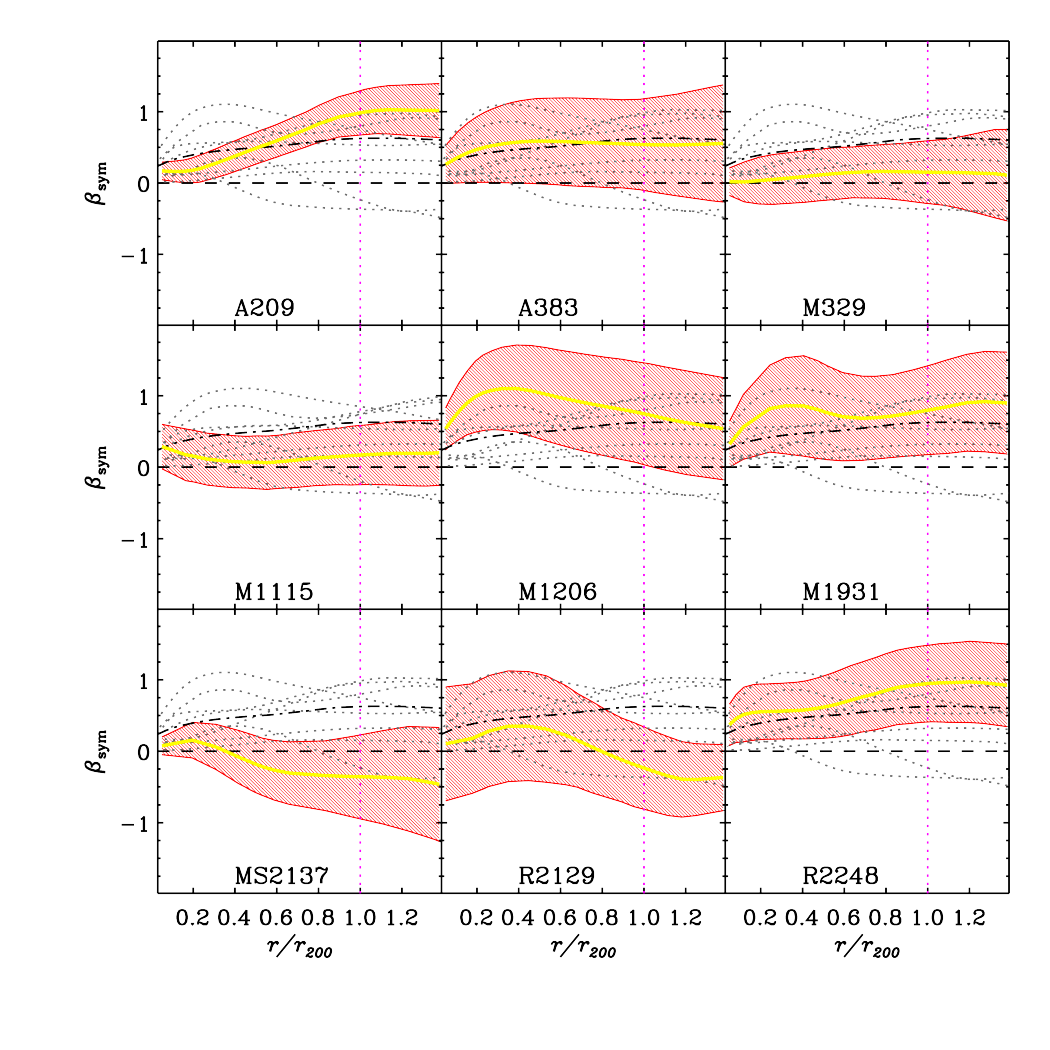}
\caption{Each panel shows the 1$\sigma$ confidence levels of a cluster \jei\ $\bs(r)$, as obtained from the procedure described in Sect.~\ref{ss:jeans} (red shadings), and its central value (yellow line), the weighted mean of the nine profiles, $\langle \bs(r) \rangle$, as defined in eq.~\ref{e:wmean} (black dash-dotted line), and the $\bs(r)$ of the other eight clusters (black dotted lines). The horizontal dashed black line indicates isotropic orbits. Orbits are radial  above  this line and  tangential below it.}
\label{f:betajei}
\end{figure*}

In Fig.~\ref{f:vdp1} we show the observed VDP (dots with error bars) and the VDP implied by the \jei\ procedure for M1206 (and for all nine clusters in Fig.~\ref{f:vdp}). By construction, the \jei\ solution for the VDP coincides with the \texttt{LOWESS} smoothed VDP, only if a dynamical equilibrium solution exists for the adopted \mr. This is indeed the case for all  nine of our clusters, and it suggests that the not-so-good agreement between the \mamp\ and the observed VDPs for some clusters was not due to a violation of dynamical equilibrium, but to the rigidity of the adopted $\br$ models. Nonetheless, the \mamp\ $\br$ are in agreement with the \jei~ $\br$ within the uncertainties (see Fig.~\ref{f:betajeimamp}), with the exception of the central region of A209. This difference might be due to the fact that the $N(R)$ models adopted in the \texttt{MAMPOSSt} analysis are not general enough to always fit the observed profiles very closely. It is perhaps worth noting that A209 is the only cluster for which the NFW model does not provide a good fit to the observed $N(R)$.

In Fig.~\ref{f:betajei} we show the main result of this article, that is, the nine individual $\bs(r)$ obtained with \jei\ (yellow lines) and the N$_{{\rm m}}$-weighted mean of these profiles (black dash-dotted line), 
\begin{equation}
\langle \bs(r) \rangle \equiv \frac{\sum_{i=1}^9 \beta_{{\rm sym,}i} N_{{\rm m,}i}} {\sum_{i=1}^9 N_{{\rm m,}i}} \label{e:wmean}.
\end{equation}
The red shadings indicate the 1$\sigma$ uncertainties on the individual cluster $\br$ obtained from the bootstrap analysis described in Sect.~\ref{ss:jeans}. 

Despite the large uncertainties, there are significant differences among the different cluster $\br$ profiles.
This indicates that there is no universal $\br$ for galaxy clusters. In this respect, the average  $\langle \beta(r) \rangle$ is not representative of the whole cluster population and we use it just as a reference. In the following, we investigate the possible causes of the different $\br$.

\subsection{Correlation with other cluster properties}\label{ss:corr}
We define the deviation of the individual cluster $\bs(r)$ from their weighted average as 
\begin{equation}
\label{e:d1}
{\rm d}\beta \equiv \sum_{i=1}^{50} \, \bs(r_i) - \langle \bs(r_i) \rangle,
\end{equation}
where the $\bs$ profiles are evaluated in 50 equally spaced radial points in the radial range we considered for our analysis (0.05 Mpc to 1.36 $r_{200}$). We note that the average profile is not representative of the whole cluster population $\beta(r)$ due to the significant variance from cluster to cluster. Here we use it as a convenient reference profile to quantify the cluster-to-cluster $\beta(r)$ variations.

\begin{table}
\centering
\caption{Quantities for regression analysis.}
\label{t:corr}
\begin{tabular}{lrrccccc} 
\toprule
Short name & ${\rm d}\beta$ & $\mvir$ & $\cvir$ & $\cnu$ & $f_{\rm blue}$ & $\fst$ \\
\midrule
      A209 & $  6$   & $17.3$    & $3.5$    & $3.5$    & $0.33$   & $0.40$  \\
      A383 & $  0$ &   $8.4$    & $7.5$    & $3.9$    & $0.46$  & $0.20$       \\
      M329 & $-19$     & $11.5$    & $5.5$    & $2.6$    & $0.51$    & $0.00$     \\
     M1115 & $-17$      & $10.7$    & $2.6$    & $1.8$    & $0.34$    & $0.31$     \\
     M1206 & $ 14$      & $15.9$    & $8.1$    & $3.1$    & $0.30$    & $0.22$     \\
     M1931 & $ 11$      & $11.5$    & $7.8$    & $1.7$    & $0.22$    & $0.10$     \\
    MS2137 & $-33$      & $ 7.9$    & $3.7$    & $1.6$    & $0.44$    & $0.47$     \\
     R2129 & $-23$      & $ 7.7$    & $3.8$    & $1.7$    & $0.29$    & $0.47$    \\
     R2248 & $ 10$      & $22.7$    & $3.2$    & $2.5$    & $0.32$    & $0.50$     \\
\bottomrule
\end{tabular}
\tablefoot{Error bars are not listed as they are not used in the regression analysis. Masses $\mvir$ are in units of $10^{14} \, M_{\odot}$.
}
\end{table}

We searched for correlations of \db with the following cluster properties to try to understand the physical reason for the substantial variance in the cluster $\bs(r)$:
\begin{itemize}
\item $\mvir$;
\item $\cvir$;
\item $\cnu \equiv \rvir/\rnu$;
\item the total fraction of blue galaxies, $f_{\rm blue}$;
\item the total fraction of galaxies in subclusters, $\fst$.
\end{itemize}
All quantities are listed in Table~\ref{t:corr}. The quantities $\mvir, \cvir, \cnu$ are the best-fit values obtained in the \mamp\ analysis. The total fraction of blue galaxies $f_b$ is taken from Maraboli et al. (in prep.). To determine $\fst$, we ran the \dsp\ subcluster analysis in its no-overlapping mode \citep{BBA23}. \dsp\ is a development of the method originally proposed by \citet{DS88}. It examines possible subclusters of different multiplicities around each cluster member, and identifies those with significant different kinematics from the cluster. We were not interested in identifying very small subclusters, but only those that are expected to leave a significant effect on the internal cluster dynamics. We therefore only looked for subclusters richer than N$_{{\rm m}}/10$. To speed up calculations, we did not consider subclusters of all possible multiplicities as in the original \dsp\ code, but only those with any of six multiplicities equally spaced between N$_{{\rm m}}/10$ and N$_{{\rm m}}/3$. We ran 1000 MonteCarlo resamplings to establish the subcluster probabilities, and retained only those with a formal probability $<1/1000$. In Fig.~\ref{f:subs} we show the spatial distributions of the subclusters found in each of our nine clusters.

\begin{figure}
\centering
\includegraphics[width=\hsize]{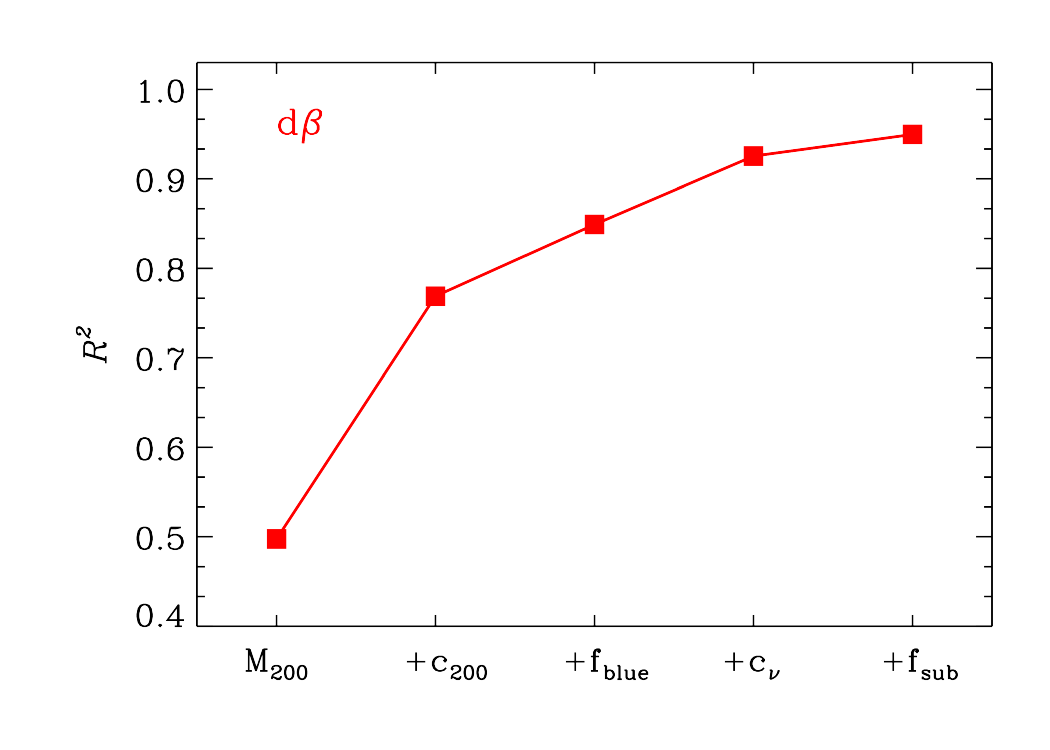}
\caption{Results of the stepwise regression analysis (forward approach). Each square is the value of the coefficient of determination, $R^2$, of ${\rm d}\beta$ vs. the quantities labeled on the x-axis, with the inclusion of an additional quantity at each new point from left to right. }
\label{f:stepwise}
\end{figure}

\begin{figure}
\centering
\includegraphics[width=\hsize]{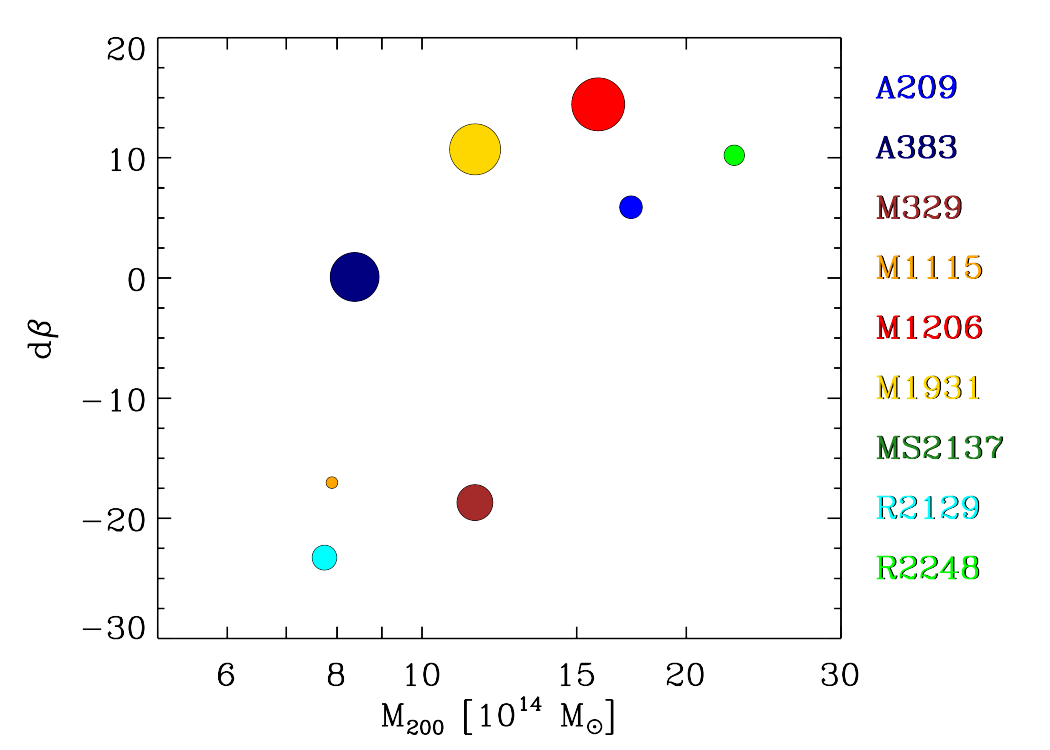}
\caption{${\rm d}\beta$ vs. $\mvir$. The symbol size is proportional to $\cvir$.
The colors identify our nine clusters, in order of increasing redshift with increasing color wavelength (from blue to brown).}
\label{f:dbetaM}
\end{figure}

We performed a stepwise regression analysis to identify   the main predictive cluster properties for ${\rm d}\beta$, using a forward selection approach \citep[][see also \citeauthor{Biviano+92b} \citeyear{Biviano+92b}, for an astrophysical application of the method]{Efroymson60}. In this procedure, we start by considering the cluster property that is most strongly correlated with ${\rm d}\beta$ and then we continue by adding the variable that contributes the largest increase in the coefficient of determination, $R^2$.\footnote{The coefficient of determination,
$R^2$, measures the proportion of variation in the dependent variable that is accounted for by the independent variables in the model. It is the square of the multiple correlation coefficient between the dependent variable (${\rm d}\beta$ in our case) and the independent variables (the cluster properties listed in Table~\ref{t:corr} in our case); see, e.g., Eq.~(10.5.25) in \citet{DeGroot87}.} In Fig.~\ref{f:stepwise} we show {$R^2$} versus the cluster properties. There are two main predictive variables for ${\rm d}\beta$, namely $\mvir$ and $\cvir$; other cluster properties contribute very little to the improvement of the regression. 

A visual confirmation of the results of the stepwise regression analysis is given in Fig.~\ref{f:dbetaM} where we show ${\rm d}\beta$ versus $\mvir$. The symbol sizes are proportional to $\cvir$. There is a clear dependence of ${\rm d}\beta$ on $\mvir$, and also on $\cvir$ at fixed $\mvir$. Clusters of higher mass, and of higher concentration at a given mass, are characterized by more radial orbits.

\subsection{Comparison with previous results}\label{ss:prev}

\begin{figure}
\centering
\includegraphics[width=\hsize]{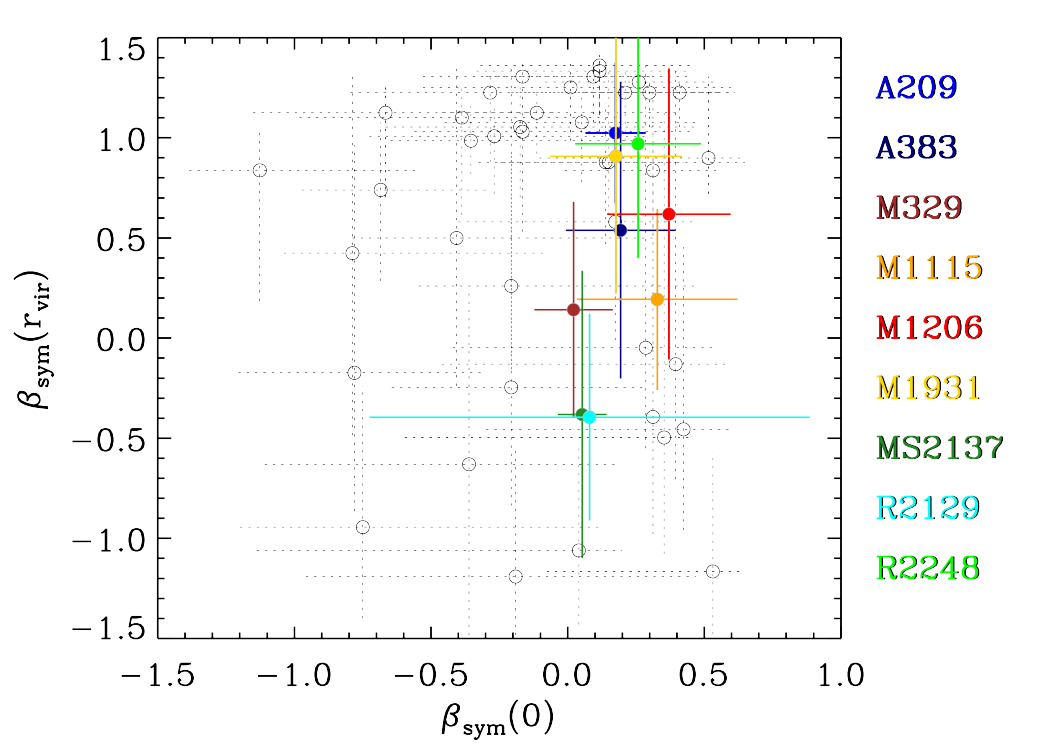}
\caption{Velocity anisotropy {$\beta_{{\rm sym}}$} at {$r=r_{{\rm vir}}$,} vs. velocity anisotropy at $r=0$, for our nine clusters (colored dots) and for the lower-$z$ clusters of \citet[black circles]{WL10}. The 1$\sigma$ error bars are shown. {The virial radius is $r_{{\rm vir}}=r_{100}$ for the nearby clusters of \citet{WL10}, and
it is approximated by $r_{126}$ for our clusters.}
The colors identify our nine clusters, in order of increasing redshift with increasing color wavelength (from blue to brown).
}
\label{f:wl10}
\end{figure}

\begin{figure}
\centering
\includegraphics[width=\hsize]{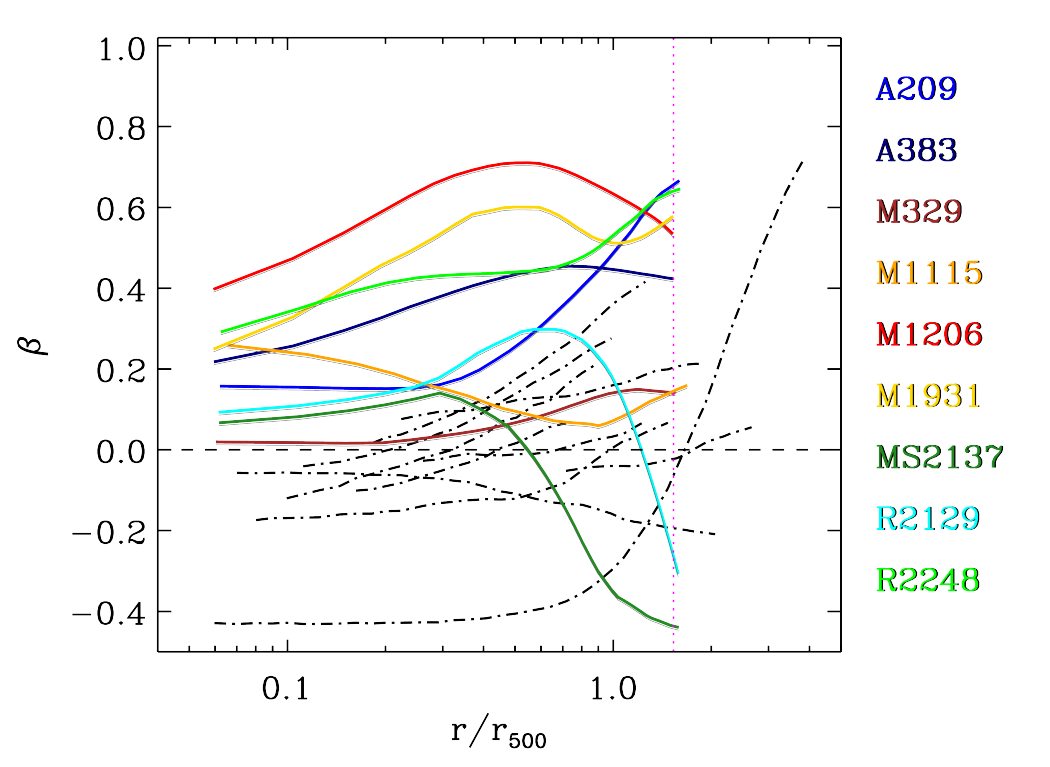}
\caption{$\br$ of our nine clusters (colored solid lines) and of the ten nearby clusters of \citet[black dot-dashed lines]{Li+23} displayed in the same format as that of Fig.~4 in \citet{Li+23}. The colors identify our nine clusters, in order of increasing redshift with increasing color wavelength (from blue to brown). The vertical magenta dotted line indicates $1.5 \, r_{500}$, which  is on average equivalent to $\rvir$ for our nine clusters.}
\label{f:li23}
\end{figure}

Here we compare our results with those of \citet{WL10} and \citet{Li+23}, who have both homogeneously measured $\br$ for several clusters. Both studies investigated low-$z$ cluster samples. The {\citet{WL10} results were based on modeling the cluster's energy and angular momentum by distribution function models. \citet{Li+23} solved the Jeans equation using ten free parameters, of which four define the model for $\br=\beta_0+(\beta_{\infty}-\beta_0) \, [1+(r_0/r)^n]^{-1}$. To better constrain these parameters, they also used the fourth moment of the los velocity distribution of cluster galaxies.}

In Fig.~\ref{f:wl10} we show {$\bs(r_{\rm{vir}})$}\footnote{{We take $r_{\rm{vir}}=r_{100}$ for the nearby cluster sample and $r_{\rm{vir}}=r_{126}$ for our clusters, following \citet{BN98}.}}  versus the values of $\bs$ estimated at the closest radius to the center of the cluster, $\bs(0)$ for simplicity, for our nine clusters and for the 41 low-$z$ clusters studied by \citet[we converted the values of $\beta$ in their Table 2 to $\bs$]{WL10}. The variance in the values of {$\bs(r_{\rm{vir}})$} is similar for our sample of clusters and the \citet{WL10} sample. However, the variance in the values of $\bs(0)$ is smaller for our sample since all the clusters in our sample have $\bs(0)$ close to isotropic or slightly radial. In contrast, many of the \citet{WL10} clusters have tangential orbits near the center. 

In Fig.~\ref{f:li23} we show our nine clusters $\br$ and their weighted average using the same format and axis ranges of Fig.~4 in \citet{Li+23}, to allow  a direct comparison with their ten low-$z$ clusters $\br$, which we reproduce in our figure. We note that the radii on the $x$-axis are in units of $r_{500}$, which we derive using the $\rvir, \rs$ values listed in Table~\ref{t:mamp}, by adopting an NFW mass density profile.
The $\br$ variance across the clusters in the sample of \citet{Li+23} and in our sample appear similar, but the clusters of \citet{Li+23} have tangential orbits near the center, and a lower degree of radial anisotropy at large radii, compared to our nine clusters.

\subsection{Comparison with simulations}\label{ss:simu}
We here compare the $\bs(r)$ profiles of our nine clusters to those of halos of similar mass and redshifts from cosmological numerical simulations. Specifically, we consider the semi-analytical model \gaea\ (already used and described in Sect.~\ref{ss:members}) and the hydrodynamical cosmological simulation of \citet[][\rf\ hereafter]{Ragone+18}. The \rf\ simulations have been shown to reproduce well the observational properties of the BCGs, both for  their evolution in mass and for their alignment with the host cluster \citep{Ragone+18,Ragone+20}.

\begin{figure}
\centering
\includegraphics[width=\hsize]{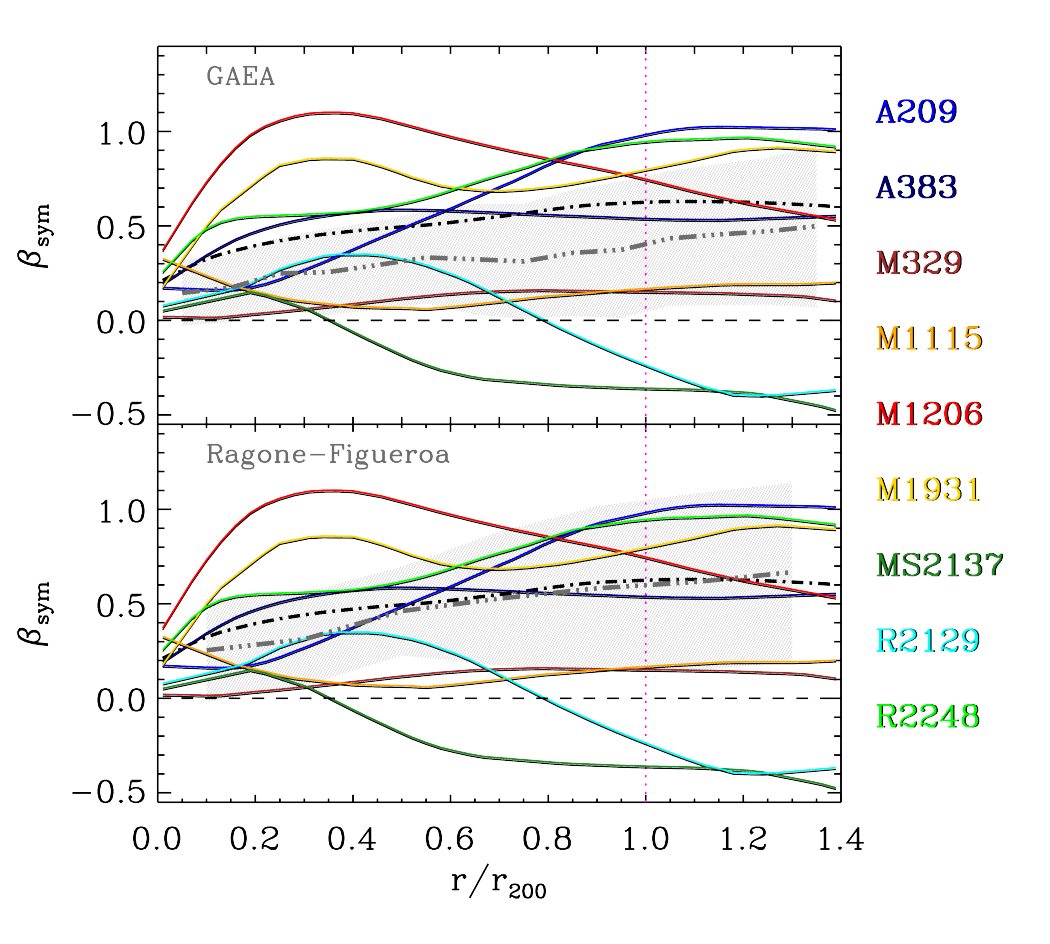}
\caption{Upper panel:
Nine clusters $\bs(r)$ (color-coded  in order of increasing redshift with increasing color wavelength, from blue to brown)
and their weighted mean velocity anisotropy profile $\langle \br \rangle$ (dash-dotted black line) compared to the mean profile of 112 halos in the \gaea\ simulation
(triple-dot-dashed gray line) and its rms (gray shading).  Lower panel: Same as the upper panel, but the comparison is made with the mean profile of 25 halos in the \rf\ simulation.}
\label{f:sims}
\end{figure}

We selected 112 halos in the redshift and mass range spanned by our nine clusters, $0.19 \leq z \leq 0.45$, $\mvir \geq 10^{14.85} \, M_{\odot}$, in \gaea, and 25 halos at $z=0.3$, and with $\mvir \geq 10^{14.85} \, M_{\odot}$, in \rf. In Fig.~\ref{f:sims} we show the $\bs(r)$ profiles of our nine clusters and their weighted mean, and the mean $\langle \bs(r) \rangle$ and its rms of the selected halos in the \gaea\ (upper panel) and \rf\ (lower panel) simulations. The \gaea\ $\langle \bs(r) \rangle$ lies below the observational mean profile, even if the difference is not statistically significant. On the other hand, the agreement between the \rf\ and observational $\langle \bs(r) \rangle$ is excellent. The variance in the halo $\bs(r)$ is somewhat larger for the observational sample than for the simulated halos, both in \gaea\ and \rf.

The $\bs(r)$ of the simulated halos are direct measurements; we have not tried to mimic the whole observational procedure. The good agreement between the observed and simulated {average} profiles, if not a pure coincidence, can be considered indirect evidence supporting the reliability of our observational results and the lack of significant systematics in our analysis. 

While this paper was being prepared for submission, we became aware of the results of \citet{Abdullah+25}, who analyzed $\br$ for halos with $\log \mvir \geq 14.05$ up to $z=1.5$ in the Uchuu-UM mock catalog of \citet{Aung+23} derived from the Uchuu cosmological N-body simulation \citep{Ishiyama+21}. Their $\langle \br \rangle$ (Fig.~2 in their paper) is completely consistent with our result. 

\section{Discussion}\label{s:disc}
By our joint lensing, \mamp, and \jei~ analysis of our sample of nine clusters, we find that the weighted average $\br$ 
is mildly radially anisotropic at all radii. The radial anisotropy increases with radius up to $\rvir$, where it reaches a plateau. There is a significant variance in $\br$ from cluster to cluster, even though the nine clusters span a narrow range in mass ($\mvir \geq 10^{14.85} \, M_{\odot}$) and redshift ($0.19 \leq z \leq 0.45$). 

A comparison of the average $\br$ of our nine clusters with those of halos of similar mass and redshifts from cosmological numerical simulations shows good agreement with the \gaea\ simulation, and excellent agreement with the \rf\ simulation. However, the $\br$ variance appears to be somewhat larger in the observational sample than in the simulated ones. It is possible that the observational variance is inflated by the 
fact that different clusters are observed at different orientations relative to their major axis, while the $\br$ of the simulated halos are averaged over all orientations. We plan to address this point in a future work.

Comparison with previous results in the literature by \citet{WL10}
shows that the $\br$ variance is higher in nearby clusters than in our cluster sample. This is caused by many nearby clusters showing more tangential orbits near the center than our clusters, as also seen in the sample of nearby clusters of \citet{Li+23}. \citet{BP09} were the first to suggest a decreasing radial anisotropy of cluster galaxy orbits with decreasing $z$. They attributed this orbital evolution to the change in the gravitational potential of the cluster due to the secular growth of the mass of the cluster via hierarchical accretion \citep{Gill+04} and/or to the mixing of energy and angular momentum that follows episodes of major mergers \citep{Valluri+07,LC11}, analogous to the violent relaxation process \citep{LyndenBell67}. An additional physical mechanism that can explain the reduction of radial anisotropy near the cluster center with time is dynamical friction \citep{Chandrasekhar43}. Even if the timescale for dynamical friction is long for individual galaxies, this process can still be effective for infalling groups \citep{Mamon+19}. The $z$-evolution of $\br$, from more radial to more isotropic orbits with time, is also seen in the Uchuu simulated halos \citep[][see their Fig. 4]{Abdullah+25}.

We tried to identify  the main cluster property that could be responsible for the $\br$ variance. We found that higher-$\mvir$ clusters, as well as clusters with high $\cvir$ at given $\mvir$, tend to have more radial orbits. Recently, \citet{Pizzuti+25} have found a similar (but not statistically significant) trend of $\beta(\rvir)$ with cluster mass by performing a stacking analysis of 75 $z<0.6$ clusters from the CHEX-MATE sample \citep{CM21}. 

A trend of increasing $\beta(\rvir)$ with halo mass {has been observed in cosmological simulations by \citet[][see their Fig. 10]{Munari+13} and by} \citet[][see their Fig. 3]{Abdullah+25}.
There is also a trend of the same nature in the \rf\ simulation, with $\bs(\rvir) \approx 0$ for halos with $\log \mvir/M_{\odot} < 14.3$, and $\approx 0.7$ for halos with $\log \mvir/M_{\odot} \geq 14.85$. \citet{Abdullah+25} attributes the mass dependence of $\br$ to the longer dynamical timescales of more massive halos, requiring more time to reach the more mature dynamical status represented by isotropic orbits. {Another possibility is that more massive clusters have a larger fraction of recently accreted galaxies, but we fail to see a significant dependence of $\br$ on the fraction of blue galaxies in our sample.
On the other hand, orbital isotropization could be caused by major mergers, that induce} rapid changes in the cluster gravitational potential, and can temporarily disrupt the central mass distribution, lowering the cluster mass concentration \citep[see, e.g.,][]{Okabe+19,Gianfagna+23}. 
 
In the future, we plan to further investigate this issue, and, more in general, the origin of the variance in cluster $\br$ profiles,
by following the evolution of the velocity anisotropy profiles of simulated halos in time (Ragone-Figueroa et al. in prep.).

\section{Summary and conclusions}\label{s:conc}
We analyzed a sample of nine massive clusters from the CLASH-VLT sample, with $\mvir \geq 10^{14.85} \, M_{\odot}$ and $0.19 \leq z \leq 0.45$, to determine their velocity anisotropy profiles from the center to {slightly beyond the virial radius}. We determined the membership of cluster galaxies with an algorithm calibrated on halos from a cosmological simulation. We then
used \mamp\ to solve the Jeans equation of dynamical equilibrium for the projected phase-space distributions of cluster members. In the \mamp\ analysis we adopted the NFW model for $M(r)$, and for eight of our nine clusters, we adopted Gaussian priors on the $\rvir, \rs$ parameters, derived from the gravitational lensing analysis of \citetalias{Umetsu+18}. We then used the \mamp\ best fit $M(r)$ in the inversion of the Jeans equation {(\jei)} to determine $\br$ of each individual cluster, without any assumption on the functional form of $\br$. 

We find that the average $\langle \br \rangle$ of our nine clusters, weighted on the number of cluster members, is radially anisotropic, increasing from $\beta \simeq 0.2$ ($\bs \simeq 0.25$) at the center to $\beta \simeq 0.5$ ($\bs \simeq 0.66$) at $r_{200}$, and flattening thereafter. The nine cluster $\br$ are not all consistent with $\langle \br \rangle$ within their 1$\sigma$ uncertainties; that is, we detect significance variance in $\br$ across the cluster sample. We find that clusters of high mass, and those with a high concentration per given mass, have more radial orbits for their member galaxies. The trend of orbits that are more radial in more massive clusters is also seen in cosmological simulations \citep{Munari+13,Abdullah+25}, and it is attributed to the longer dynamical timescales of more massive halos. The trend of orbits that are less radial in clusters of lower concentrations could instead be due to major mergers that can cause a decrease in concentration \citep{Okabe+19,Gianfagna+23} and, at the same time, lead to the isotropization of the galaxy orbits \citep{Valluri+07}.

A comparison with results from the literature \citep{WL10,Li+23}, show that clusters at lower $z$ have higher $\br$ variance near the center, caused by many clusters in the low-$z$ samples displaying tangential orbits, unlike the nine clusters in our sample that are characterized by mildly radial or isotropic orbits near the center. This orbital evolution with $z$ was already suggested by \citet{BP09}. We propose three physical mechanisms that can possibly lead to a reduction in the radial anisotropy of cluster galaxy orbits: (i) dynamical friction \citep{Chandrasekhar43,Mamon+19}, (ii) the evolution of the cluster gravitational potential due to mass accretion \citep{Gill+04}, and (iii) violent relaxation following major merger events \citep{Valluri+07}. 

We also compared our $\br$ to those of halos of similar mass and $z$, in two cosmological simulations. {There is a good} agreement of the average $\br$ profile with {those of the simulated halos, in particular the one from} the hydrodynamical simulations of \citet{Ragone+18}. In both simulations, the $\br$ variance is slightly smaller than the observed one. In the future we plan to investigate the $\br$ profiles of cluster-sized halos in hydrodynamical simulations in order to uncover the physical mechanisms affecting the orbits of cluster galaxies (Ragone-Figueroa et al. in prep.).

\begin{acknowledgements}
We thank the referee for her/his careful, competent, and useful report, and Gary Mamon and 
Massimiliano Parente for useful discussion. We acknowledge financial support through the grants PRIN-MIUR 2017WSCC32 and 2020SKSTHZ. AB acknowledges the financial contribution from the INAF mini-grant 1.05.12.04.01 "The dynamics of clusters of galaxies from the projected phase-space distribution of cluster galaxies". We acknowledge financial support from the European Union's HORIZON-MSCA-2021-SE-01 Research and Innovation Programme under the Marie Sklodowska-Curie grant agreement number 101086388 - Project (LACEGAL).\end{acknowledgements}

\bibliographystyle{aa}
\bibliography{master}

\begin{thebibliography}{96}
\expandafter\ifx\csname natexlab\endcsname\relax\def\natexlab#1{#1}\fi

\bibitem[{{Abdullah} {et~al.}(2025){Abdullah}, {Mabrouk}, {Ishiyama}, {Wilson},
  {Amin}, {Khattab}, \& {Abdel Rahman}}]{Abdullah+25}
{Abdullah}, M.~H., {Mabrouk}, R.~H., {Ishiyama}, T., {et~al.} 2025, \apj, 987,
  70

\bibitem[{{Adami} {et~al.}(2009){Adami}, {Le Brun}, {Biviano}, {Durret},
  {Lamareille}, {Pell{\'o}}, {Ilbert}, {Mazure}, {Trilling}, \&
  {Ulmer}}]{Adami+09}
{Adami}, C., {Le Brun}, V., {Biviano}, A., {et~al.} 2009, \aap, 507, 1225

\bibitem[{{Aguerri} {et~al.}(2017){Aguerri}, {Agulli}, {Diaferio}, \& {Dalla
  Vecchia}}]{AADDV17}
{Aguerri}, J.~A.~L., {Agulli}, I., {Diaferio}, A., \& {Dalla Vecchia}, C. 2017,
  \mnras, 468, 364

\bibitem[{{Aguirre Tagliaferro} {et~al.}(2021){Aguirre Tagliaferro}, {Biviano},
  {De Lucia}, {Munari}, \& {Garcia Lambas}}]{Tagliaferro+21}
{Aguirre Tagliaferro}, T., {Biviano}, A., {De Lucia}, G., {Munari}, E., \&
  {Garcia Lambas}, D. 2021, \aap, 652, A90

\bibitem[{{Annunziatella} {et~al.}(2016){Annunziatella}, {Mercurio}, {Biviano},
  {Girardi}, {Nonino}, {Balestra}, {Rosati}, {Bartosch Caminha}, {Brescia},
  {Gobat}, {Grillo}, {Lombardi}, {Sartoris}, {De Lucia}, {Demarco}, {Frye},
  {Fritz}, {Moustakas}, {Scodeggio}, {Kuchner}, {Maier}, \&
  {Ziegler}}]{Annunziatella+16}
{Annunziatella}, M., {Mercurio}, A., {Biviano}, A., {et~al.} 2016, \aap, 585,
  A160

\bibitem[{{Aung} {et~al.}(2023){Aung}, {Nagai}, {Klypin}, {Behroozi},
  {Abdullah}, {Ishiyama}, {Prada}, {P{\'e}rez}, {L{\'o}pez Cacheiro}, \&
  {Ruedas}}]{Aung+23}
{Aung}, H., {Nagai}, D., {Klypin}, A., {et~al.} 2023, \mnras, 519, 1648

\bibitem[{{Bacon} {et~al.}(2010){Bacon}, {Accardo}, {Adjali}, {Anwand},
  {Bauer}, {Biswas}, {Blaizot}, {Boudon}, {Brau-Nogue}, {Brinchmann},
  {Caillier}, {Capoani}, {Carollo}, {Contini}, {Couderc}, {Daguis{\'e}},
  {Deiries}, {Delabre}, {Dreizler}, {Dubois}, {Dupieux}, {Dupuy}, {Emsellem},
  {Fechner}, {Fleischmann}, {Fran{\c{c}}ois}, {Gallou}, {Gharsa}, {Glindemann},
  {Gojak}, {Guiderdoni}, {Hansali}, {Hahn}, {Jarno}, {Kelz}, {Koehler},
  {Kosmalski}, {Laurent}, {Le Floch}, {Lilly}, {Lizon}, {Loupias}, {Manescau},
  {Monstein}, {Nicklas}, {Olaya}, {Pares}, {Pasquini}, {P{\'e}contal-Rousset},
  {Pell{\'o}}, {Petit}, {Popow}, {Reiss}, {Remillieux}, {Renault}, {Roth},
  {Rupprecht}, {Serre}, {Schaye}, {Soucail}, {Steinmetz}, {Streicher}, {Stuik},
  {Valentin}, {Vernet}, {Weilbacher}, {Wisotzki}, \& {Yerle}}]{Bacon+10}
{Bacon}, R., {Accardo}, M., {Adjali}, L., {et~al.} 2010, in Society of
  Photo-Optical Instrumentation Engineers (SPIE) Conference Series, Vol. 7735,
  Ground-based and Airborne Instrumentation for Astronomy III, ed. I.~S.
  {McLean}, S.~K. {Ramsay}, \& H.~{Takami}, 773508

\bibitem[{{Balestra} {et~al.}(2016){Balestra}, {Mercurio}, {Sartoris},
  {Girardi}, {Grillo}, {Nonino}, {Rosati}, {Biviano}, {Ettori}, {Forman},
  {Jones}, {Koekemoer}, {Medezinski}, {Merten}, {Ogrean}, {Tozzi}, {Umetsu},
  {Vanzella}, {van Weeren}, {Zitrin}, {Annunziatella}, {Caminha}, {Broadhurst},
  {Coe}, {Donahue}, {Fritz}, {Frye}, {Kelson}, {Lombardi}, {Maier},
  {Meneghetti}, {Monna}, {Postman}, {Scodeggio}, {Seitz}, \&
  {Ziegler}}]{Balestra+16}
{Balestra}, I., {Mercurio}, A., {Sartoris}, B., {et~al.} 2016, \apjs, 224, 33

\bibitem[{{Bartalesi} {et~al.}(2025){Bartalesi}, {Ettori}, \& {Nipoti}}]{BEN25}
{Bartalesi}, T., {Ettori}, S., \& {Nipoti}, C. 2025, \aap, 697, A17

\bibitem[{{Bartelmann}(1996)}]{Bartelmann96}
{Bartelmann}, M. 1996, \aap, 313, 697

\bibitem[{{Battaglia} {et~al.}(2008){Battaglia}, {Helmi}, {Tolstoy}, {Irwin},
  {Hill}, \& {Jablonka}}]{Battaglia+08}
{Battaglia}, G., {Helmi}, A., {Tolstoy}, E., {et~al.} 2008, \apjl, 681, L13

\bibitem[{{Benatov} {et~al.}(2006){Benatov}, {Rines}, {Natarajan}, {Kravtsov},
  \& {Nagai}}]{Benatov+06}
{Benatov}, L., {Rines}, K., {Natarajan}, P., {Kravtsov}, A., \& {Nagai}, D.
  2006, \mnras, 370, 427

\bibitem[{{Benavides} {et~al.}(2023){Benavides}, {Biviano}, \& {Abadi}}]{BBA23}
{Benavides}, J.~A., {Biviano}, A., \& {Abadi}, M.~G. 2023, \aap, 669, A147

\bibitem[{{Binney} \& {Mamon}(1982)}]{BM82}
{Binney}, J. \& {Mamon}, G.~A. 1982, \mnras, 200, 361

\bibitem[{{Binney} \& {Tremaine}(1987)}]{BT87}
{Binney}, J. \& {Tremaine}, S. 1987, Galactic dynamics (Princeton, NJ,
  Princeton University Press, 1987, 747 p.)

\bibitem[{{Biviano} {et~al.}(1992){Biviano}, {Girardi}, {Giuricin},
  {Mardirossian}, \& {Mezzetti}}]{Biviano+92b}
{Biviano}, A., {Girardi}, M., {Giuricin}, G., {Mardirossian}, F., \&
  {Mezzetti}, M. 1992, in ASSL Vol. 178: Morphological and Physical
  Classification of Galaxies, ed. G.~{Longo}, M.~{Capaccioli}, \&
  G.~{Busarello}, 413--+

\bibitem[{{Biviano} \& {Katgert}(2003)}]{BK03}
{Biviano}, A. \& {Katgert}, P. 2003, \apss, 285, 25

\bibitem[{{Biviano} \& {Katgert}(2004)}]{BK04}
{Biviano}, A. \& {Katgert}, P. 2004, \aap, 424, 779

\bibitem[{{Biviano} {et~al.}(2017){Biviano}, {Moretti}, {Paccagnella},
  {Poggianti}, {Bettoni}, {Gullieuszik}, {Vulcani}, {Fasano}, {D'Onofrio},
  {Fritz}, \& {Cava}}]{Biviano+17a}
{Biviano}, A., {Moretti}, A., {Paccagnella}, A., {et~al.} 2017, \aap, 607, A81

\bibitem[{{Biviano} {et~al.}(2023){Biviano}, {Pizzuti}, {Mercurio}, {Sartoris},
  {Rosati}, {Ettori}, {Girardi}, {Grillo}, {Caminha}, \& {Nonino}}]{Biviano+23}
{Biviano}, A., {Pizzuti}, L., {Mercurio}, A., {et~al.} 2023, \apj, 958, 148

\bibitem[{{Biviano} \& {Poggianti}(2009)}]{BP09}
{Biviano}, A. \& {Poggianti}, B.~M. 2009, \aap, 501, 419

\bibitem[{{Biviano} {et~al.}(2024){Biviano}, {Poggianti}, {Jaff{\'e}},
  {Louren{\c{c}}o}, {Pizzuti}, {Moretti}, \& {Vulcani}}]{Biviano+24}
{Biviano}, A., {Poggianti}, B.~M., {Jaff{\'e}}, Y., {et~al.} 2024, \apj, 965,
  117

\bibitem[{{Biviano} {et~al.}(2013){Biviano}, {Rosati}, {Balestra}, {Mercurio},
  {Girardi}, {Nonino}, {Grillo}, {Scodeggio}, {Lemze}, {Kelson}, {Umetsu},
  {Postman}, {Zitrin}, {Czoske}, {Ettori}, {Fritz}, {Lombardi}, {Maier},
  {Medezinski}, {Mei}, {Presotto}, {Strazzullo}, {Tozzi}, {Ziegler},
  {Annunziatella}, {Bartelmann}, {Benitez}, {Bradley}, {Brescia}, {Broadhurst},
  {Coe}, {Demarco}, {Donahue}, {Ford}, {Gobat}, {Graves}, {Koekemoer},
  {Kuchner}, {Melchior}, {Meneghetti}, {Merten}, {Moustakas}, {Munari}, {Reg{\H
  o}s}, {Sartoris}, {Seitz}, \& {Zheng}}]{Biviano+13}
{Biviano}, A., {Rosati}, P., {Balestra}, I., {et~al.} 2013, \aap, 558, A1

\bibitem[{{Biviano} \& {Salucci}(2006)}]{BS06}
{Biviano}, A. \& {Salucci}, P. 2006, \aap, 452, 75

\bibitem[{{Biviano} {et~al.}(2021){Biviano}, {van der Burg}, {Balogh},
  {Munari}, {Cooper}, {De Lucia}, {Demarco}, {Jablonka}, {Muzzin}, {Nantais},
  {Old}, {Rudnick}, {Vulcani}, {Wilson}, {Yee}, {Zaritsky}, {Cerulo}, {Chan},
  {Finoguenov}, {Gilbank}, {Lidman}, {Pintos-Castro}, \&
  {Shipley}}]{Biviano+21}
{Biviano}, A., {van der Burg}, R.~F.~J., {Balogh}, M.~L., {et~al.} 2021, \aap,
  650, A105

\bibitem[{{Biviano} {et~al.}(2016){Biviano}, {van der Burg}, {Muzzin},
  {Sartoris}, {Wilson}, \& {Yee}}]{Biviano+16}
{Biviano}, A., {van der Burg}, R.~F.~J., {Muzzin}, A., {et~al.} 2016, \aap,
  594, A51

\bibitem[{{Bryan} \& {Norman}(1998)}]{BN98}
{Bryan}, G.~L. \& {Norman}, M.~L. 1998, \apj, 495, 80

\bibitem[{{Caminha} {et~al.}(2017){Caminha}, {Grillo}, {Rosati}, {Meneghetti},
  {Mercurio}, {Ettori}, {Balestra}, {Biviano}, {Umetsu}, {Vanzella},
  {Annunziatella}, {Bonamigo}, {Delgado-Correal}, {Girardi}, {Lombardi},
  {Nonino}, {Sartoris}, {Tozzi}, {Bartelmann}, {Bradley}, {Caputi}, {Coe},
  {Ford}, {Fritz}, {Gobat}, {Postman}, {Seitz}, \& {Zitrin}}]{Caminha+17}
{Caminha}, G.~B., {Grillo}, C., {Rosati}, P., {et~al.} 2017, \aap, 607, A93

\bibitem[{{Capasso} {et~al.}(2019){Capasso}, {Saro}, {Mohr}, {Biviano},
  {Bocquet}, {Strazzullo}, {Grandis}, {Applegate}, {Bayliss}, {Benson},
  {Bleem}, {Brodwin}, {Bulbul}, {Carlstrom}, {Chiu}, {Dietrich}, {Gupta}, {de
  Haan}, {Hlavacek-Larrondo}, {Klein}, {von der Linden}, {McDonald}, {Rapetti},
  {Reichardt}, {Sharon}, {Stalder}, {Stanford}, {Stark}, {Stern}, \&
  {Zenteno}}]{Capasso+19}
{Capasso}, R., {Saro}, A., {Mohr}, J.~J., {et~al.} 2019, \mnras, 482, 1043

\bibitem[{{Chandrasekhar}(1943)}]{Chandrasekhar43}
{Chandrasekhar}, S. 1943, \apj, 97, 255

\bibitem[{{CHEX-MATE Collaboration} {et~al.}(2021){CHEX-MATE Collaboration},
  {Arnaud}, {Ettori}, {Pratt}, {Rossetti}, {Eckert}, {Gastaldello}, {Gavazzi},
  {Kay}, {Lovisari}, {Maughan}, {Pointecouteau}, {Sereno}, {Bartalucci},
  {Bonafede}, {Bourdin}, {Cassano}, {Duffy}, {Iqbal}, {Maurogordato}, {Rasia},
  {Sayers}, {Andrade-Santos}, {Aussel}, {Barnes}, {Barrena}, {Borgani},
  {Burkutean}, {Clerc}, {Corasaniti}, {Cuillandre}, {De Grandi}, {De Petris},
  {Dolag}, {Donahue}, {Ferragamo}, {Gaspari}, {Ghizzardi}, {Gitti}, {Haines},
  {Jauzac}, {Johnston-Hollitt}, {Jones}, {K{\'e}ruzor{\'e}}, {Le Brun},
  {Mayet}, {Mazzotta}, {Melin}, {Molendi}, {Nonino}, {Okabe}, {Paltani},
  {Perotto}, {Pires}, {Radovich}, {Rubino-Martin}, {Salvati}, {Saro},
  {Sartoris}, {Schellenberger}, {Streblyanska}, {Tarr{\'\i}o}, {Tozzi},
  {Umetsu}, {van der Burg}, {Vazza}, {Venturi}, {Yepes}, \& {Zarattini}}]{CM21}
{CHEX-MATE Collaboration}, {Arnaud}, M., {Ettori}, S., {et~al.} 2021, \aap,
  650, A104

\bibitem[{{Cleveland} \& {McGill}(1984)}]{CMG84}
{Cleveland}, W. \& {McGill}, R. 1984, J. Am. Stat. Assoc., 79, 807

\bibitem[{{De Lucia} \& {Blaizot}(2007)}]{DLB07}
{De Lucia}, G. \& {Blaizot}, J. 2007, \mnras, 375, 2

\bibitem[{{De Lucia} {et~al.}(2024){De Lucia}, {Fontanot}, {Xie}, \&
  {Hirschmann}}]{DeLucia+24}
{De Lucia}, G., {Fontanot}, F., {Xie}, L., \& {Hirschmann}, M. 2024, \aap, 687,
  A68

\bibitem[{{DeGroot}(1987)}]{DeGroot87}
{DeGroot}, M.~H. 1987, Probability and Statistics - Second Edition (Reading,
  MA, USA: Addison Wesley)

\bibitem[{{Dejonghe} \& {Merritt}(1992)}]{DM92}
{Dejonghe}, H. \& {Merritt}, D. 1992, \apj, 391, 531

\bibitem[{{Donahue} {et~al.}(2014){Donahue}, {Voit}, {Mahdavi}, {Umetsu},
  {Ettori}, {Merten}, {Postman}, {Hoffer}, {Baldi}, {Coe}, {Czakon},
  {Bartelmann}, {Benitez}, {Bouwens}, {Bradley}, {Broadhurst}, {Ford},
  {Gastaldello}, {Grillo}, {Infante}, {Jouvel}, {Koekemoer}, {Kelson}, {Lahav},
  {Lemze}, {Medezinski}, {Melchior}, {Meneghetti}, {Molino}, {Moustakas},
  {Moustakas}, {Nonino}, {Rosati}, {Sayers}, {Seitz}, {Van der Wel}, {Zheng},
  \& {Zitrin}}]{Donahue+14}
{Donahue}, M., {Voit}, G.~M., {Mahdavi}, A., {et~al.} 2014, \apj, 794, 136

\bibitem[{{Dressler} \& {Shectman}(1988)}]{DS88}
{Dressler}, A. \& {Shectman}, S.~A. 1988, \aj, 95, 985

\bibitem[{{Efroymson}(1960)}]{Efroymson60}
{Efroymson}, M.~A. 1960, {Multiple regression analysis}, ed. {Ralston, A. and
  Wilf, H. S.} (Wiley, New York)

\bibitem[{{Gebhardt} {et~al.}(1994){Gebhardt}, {Pryor}, {Williams}, \&
  {Hesser}}]{Gebhardt+94}
{Gebhardt}, K., {Pryor}, C., {Williams}, T.~B., \& {Hesser}, J.~E. 1994, \aj,
  107, 2067

\bibitem[{{Geller} {et~al.}(1999){Geller}, {Diaferio}, \& {Kurtz}}]{GDK99}
{Geller}, M.~J., {Diaferio}, A., \& {Kurtz}, M.~J. 1999, \apjl, 517, L23

\bibitem[{Gelman \& Rubin(1992)}]{GR92}
Gelman, A. \& Rubin, D. 1992, Statistical Science, 7, 457

\bibitem[{{Gianfagna} {et~al.}(2023){Gianfagna}, {Rasia}, {Cui}, {De Petris},
  {Yepes}, {Contreras-Santos}, \& {Knebe}}]{Gianfagna+23}
{Gianfagna}, G., {Rasia}, E., {Cui}, W., {et~al.} 2023, \mnras, 518, 4238

\bibitem[{{Gill} {et~al.}(2004){Gill}, {Knebe}, {Gibson}, \&
  {Dopita}}]{Gill+04}
{Gill}, S.~P.~D., {Knebe}, A., {Gibson}, B.~K., \& {Dopita}, M.~A. 2004,
  \mnras, 351, 410

\bibitem[{{Girardi} {et~al.}(2024){Girardi}, {Boschin}, {Mercurio}, {Nocerino},
  {Nonino}, {Rosati}, {Biviano}, {Demarco}, {Grillo}, {Sartoris}, {Tozzi}, \&
  {Vanzella}}]{Girardi+24}
{Girardi}, M., {Boschin}, W., {Mercurio}, A., {et~al.} 2024, \aap, 692, A175

\bibitem[{{Gruen} {et~al.}(2013){Gruen}, {Brimioulle}, {Seitz}, {Lee}, {Young},
  {Koppenhoefer}, {Eichner}, {Riffeser}, {Vikram}, {Weidinger}, \&
  {Zenteno}}]{Gruen+13}
{Gruen}, D., {Brimioulle}, F., {Seitz}, S., {et~al.} 2013, MNRAS, 432, 1455

\bibitem[{{Hwang} \& {Lee}(2008)}]{HL08}
{Hwang}, H.~S. \& {Lee}, M.~G. 2008, \apj, 676, 218

\bibitem[{{Ishiyama} {et~al.}(2021){Ishiyama}, {Prada}, {Klypin}, {Sinha},
  {Metcalf}, {Jullo}, {Altieri}, {Cora}, {Croton}, {de la Torre},
  {Mill{\'a}n-Calero}, {Oogi}, {Ruedas}, \& {Vega-Mart{\'\i}nez}}]{Ishiyama+21}
{Ishiyama}, T., {Prada}, F., {Klypin}, A.~A., {et~al.} 2021, \mnras, 506, 4210

\bibitem[{{Karman} {et~al.}(2017){Karman}, {Caputi}, {Caminha}, {Gronke},
  {Grillo}, {Balestra}, {Rosati}, {Vanzella}, {Coe}, {Dijkstra}, {Koekemoer},
  {McLeod}, {Mercurio}, \& {Nonino}}]{Karman+17}
{Karman}, W., {Caputi}, K.~I., {Caminha}, G.~B., {et~al.} 2017, \aap, 599, A28

\bibitem[{{Katgert} {et~al.}(2004){Katgert}, {Biviano}, \& {Mazure}}]{KBM04}
{Katgert}, P., {Biviano}, A., \& {Mazure}, A. 2004, \apj, 600, 657

\bibitem[{{King}(1962)}]{King62_denslaw}
{King}, I. 1962, \aj, 67, 274

\bibitem[{{Lapi} \& {Cavaliere}(2011)}]{LC11}
{Lapi}, A. \& {Cavaliere}, A. 2011, \apj, 743, 127

\bibitem[{{Le F{\`e}vre} {et~al.}(2003){Le F{\`e}vre}, {Saisse}, {Mancini},
  {Brau-Nogue}, {Caputi}, {Castinel}, {D'Odorico}, {Garilli}, {Kissler-Patig},
  {Lucuix}, {Mancini}, {Pauget}, {Sciarretta}, {Scodeggio}, {Tresse}, \&
  {Vettolani}}]{LeFevre+03}
{Le F{\`e}vre}, O., {Saisse}, M., {Mancini}, D., {et~al.} 2003, in Society of
  Photo-Optical Instrumentation Engineers (SPIE) Conference Series, Vol. 4841,
  Society of Photo-Optical Instrumentation Engineers (SPIE) Conference Series,
  ed. M.~{Iye} \& A.~F.~M. {Moorwood}, 1670--1681

\bibitem[{{Li} {et~al.}(2023){Li}, {Tian}, {J{\'u}lio}, {Pawlowski}, {Lelli},
  {McGaugh}, {Schombert}, {Read}, {Yu}, \& {Ko}}]{Li+23}
{Li}, P., {Tian}, Y., {J{\'u}lio}, M.~P., {et~al.} 2023, \aap, 677, A24

\bibitem[{{Li} {et~al.}(2007){Li}, {Mo}, {van den Bosch}, \& {Lin}}]{Li+07}
{Li}, Y., {Mo}, H.~J., {van den Bosch}, F.~C., \& {Lin}, W.~P. 2007, \mnras,
  379, 689

\bibitem[{{{\L}okas} \& {Mamon}(2003)}]{LM03}
{{\L}okas}, E.~L. \& {Mamon}, G.~A. 2003, \mnras, 343, 401

\bibitem[{{Lotz} {et~al.}(2019){Lotz}, {Remus}, {Dolag}, {Biviano}, \&
  {Burkert}}]{Lotz+19}
{Lotz}, M., {Remus}, R.-S., {Dolag}, K., {Biviano}, A., \& {Burkert}, A. 2019,
  \mnras, 488, 5370

\bibitem[{{Lynden-Bell}(1967)}]{LyndenBell67}
{Lynden-Bell}, D. 1967, \mnras, 136, 101

\bibitem[{{Mamon} {et~al.}(2013){Mamon}, {Biviano}, \& {Bou{\'e}}}]{MBB13}
{Mamon}, G.~A., {Biviano}, A., \& {Bou{\'e}}, G. 2013, \mnras, 429, 3079

\bibitem[{{Mamon} {et~al.}(2010){Mamon}, {Biviano}, \& {Murante}}]{MBM10}
{Mamon}, G.~A., {Biviano}, A., \& {Murante}, G. 2010, \aap, 520, A30

\bibitem[{{Mamon} {et~al.}(2019){Mamon}, {Cava}, {Biviano}, {Moretti},
  {Poggianti}, \& {Bettoni}}]{Mamon+19}
{Mamon}, G.~A., {Cava}, A., {Biviano}, A., {et~al.} 2019, \aap, 631, A131

\bibitem[{{Manolopoulou} \& {Plionis}(2017)}]{MP17}
{Manolopoulou}, M. \& {Plionis}, M. 2017, \mnras, 465, 2616

\bibitem[{{Mercurio} {et~al.}(2021){Mercurio}, {Rosati}, {Biviano},
  {Annunziatella}, {Girardi}, {Sartoris}, {Nonino}, {Brescia}, {Riccio},
  {Grillo}, {Balestra}, {Caminha}, {De Lucia}, {Gobat}, {Seitz}, {Tozzi},
  {Scodeggio}, {Vanzella}, {Angora}, {Bergamini}, {Borgani}, {Demarco},
  {Meneghetti}, {Strazzullo}, {Tortorelli}, {Umetsu}, {Fritz}, {Gruen},
  {Kelson}, {Lombardi}, {Maier}, {Postman}, {Rodighiero}, \&
  {Ziegler}}]{Mercurio+21}
{Mercurio}, A., {Rosati}, P., {Biviano}, A., {et~al.} 2021, \aap, 656, A147

\bibitem[{{Merritt}(1985)}]{Merritt85}
{Merritt}, D. 1985, \apj, 289, 18

\bibitem[{{Merritt}(1987)}]{Merritt87}
{Merritt}, D. 1987, \apj, 313, 121

\bibitem[{{Miyazaki} {et~al.}(2002){Miyazaki}, {Komiyama}, {Sekiguchi},
  {Okamura}, {Doi}, {Furusawa}, {Hamabe}, {Imi}, {Kimura}, {Nakata}, {Okada},
  {Ouchi}, {Shimasaku}, {Yagi}, \& {Yasuda}}]{Miyazaki+02}
{Miyazaki}, S., {Komiyama}, Y., {Sekiguchi}, M., {et~al.} 2002, Publications of
  the Astronomical Society of Japan, 54, 833

\bibitem[{{Munari} {et~al.}(2013){Munari}, {Biviano}, {Borgani}, {Murante}, \&
  {Fabjan}}]{Munari+13}
{Munari}, E., {Biviano}, A., {Borgani}, S., {Murante}, G., \& {Fabjan}, D.
  2013, \mnras, 430, 2638

\bibitem[{{Munari} {et~al.}(2014){Munari}, {Biviano}, \& {Mamon}}]{MBM14}
{Munari}, E., {Biviano}, A., \& {Mamon}, G.~A. 2014, \aap, 566, A68

\bibitem[{{Natarajan} \& {Kneib}(1996)}]{NK96}
{Natarajan}, P. \& {Kneib}, J.-P. 1996, \mnras, 283, 1031

\bibitem[{{Navarro} {et~al.}(1996){Navarro}, {Frenk}, \& {White}}]{NFW96}
{Navarro}, J.~F., {Frenk}, C.~S., \& {White}, S. D.~M. 1996, \apj, 462, 563

\bibitem[{{Navarro} {et~al.}(1997){Navarro}, {Frenk}, \& {White}}]{NFW97}
{Navarro}, J.~F., {Frenk}, C.~S., \& {White}, S. D.~M. 1997, \apj, 490, 493

\bibitem[{{Okabe} {et~al.}(2019){Okabe}, {Oguri}, {Akamatsu}, {Hamabata},
  {Nishizawa}, {Medezinski}, {Koyama}, {Hayashi}, {Okabe}, {Ueda}, {Mitsuishi},
  \& {Ota}}]{Okabe+19}
{Okabe}, N., {Oguri}, M., {Akamatsu}, H., {et~al.} 2019, \pasj, 71, 79

\bibitem[{{Osipkov}(1979)}]{Osipkov79}
{Osipkov}, L.~P. 1979, Soviet Astronomy Letters, 5, 42

\bibitem[{{Pizzardo} {et~al.}(2024){Pizzardo}, {Geller}, {Kenyon}, \&
  {Damjanov}}]{PGKD24}
{Pizzardo}, M., {Geller}, M.~J., {Kenyon}, S.~J., \& {Damjanov}, I. 2024, \aap,
  683, A82

\bibitem[{{Pizzuti} {et~al.}(2025){Pizzuti}, {Barrena}, {Sereno},
  {Streblyanska}, {Ferragamo}, {Maurogordato}, {Cappi}, {Ettori}, {Pratt},
  {Castignani}, {Donahue}, {Eckert}, {Gastaldello}, {Gavazzi}, {Haines}, {Kay},
  {Lovisari}, {}, {Maughan}, {Pointecouteau}, {Rasia}, {Radovich}, \&
  {Sayers}}]{Pizzuti+25}
{Pizzuti}, L., {Barrena}, R., {Sereno}, M., {et~al.} 2025, arXiv e-prints,
  arXiv:2504.03708

\bibitem[{{Pizzuti} {et~al.}(2023){Pizzuti}, {Saltas}, {Biviano}, {Mamon}, \&
  {Amendola}}]{Pizzuti+23}
{Pizzuti}, L., {Saltas}, I., {Biviano}, A., {Mamon}, G., \& {Amendola}, L.
  2023, The Journal of Open Source Software, 8, 4800

\bibitem[{{Postman} {et~al.}(2012){Postman}, {Coe}, {Ben{\'{\i}}tez},
  {Bradley}, {Broadhurst}, {Donahue}, {Ford}, {Graur}, {Graves}, {Jouvel},
  {Koekemoer}, {Lemze}, {Medezinski}, {Molino}, {Moustakas}, {Ogaz}, {Riess},
  {Rodney}, {Rosati}, {Umetsu}, {Zheng}, {Zitrin}, {Bartelmann}, {Bouwens},
  {Czakon}, {Golwala}, {Host}, {Infante}, {Jha}, {Jimenez-Teja}, {Kelson},
  {Lahav}, {Lazkoz}, {Maoz}, {McCully}, {Melchior}, {Meneghetti}, {Merten},
  {Moustakas}, {Nonino}, {Patel}, {Reg{\"o}s}, {Sayers}, {Seitz}, \& {Van der
  Wel}}]{Postman+12}
{Postman}, M., {Coe}, D., {Ben{\'{\i}}tez}, N., {et~al.} 2012, \apjs, 199, 25

\bibitem[{{Ragone-Figueroa} {et~al.}(2020){Ragone-Figueroa}, {Granato},
  {Borgani}, {De Propris}, {Garc{\'\i}a Lambas}, {Murante}, {Rasia}, \&
  {West}}]{Ragone+20}
{Ragone-Figueroa}, C., {Granato}, G.~L., {Borgani}, S., {et~al.} 2020, \mnras,
  495, 2436

\bibitem[{{Ragone-Figueroa} {et~al.}(2018){Ragone-Figueroa}, {Granato},
  {Ferraro}, {Murante}, {Biffi}, {Borgani}, {Planelles}, \&
  {Rasia}}]{Ragone+18}
{Ragone-Figueroa}, C., {Granato}, G.~L., {Ferraro}, M.~E., {et~al.} 2018,
  \mnras, 479, 1125

\bibitem[{{Read} {et~al.}(2021){Read}, {Mamon}, {Vasiliev}, {Watkins},
  {Walker}, {Pe{\~n}arrubia}, {Wilkinson}, {Dehnen}, \& {Das}}]{Read+21}
{Read}, J.~I., {Mamon}, G.~A., {Vasiliev}, E., {et~al.} 2021, \mnras, 501, 978

\bibitem[{{Rosati} {et~al.}(2014){Rosati}, {Balestra}, {Grillo}, {Mercurio},
  {Nonino}, {Biviano}, {Girardi}, {Vanzella}, \& {Clash-VLT Team}}]{Rosati+14}
{Rosati}, P., {Balestra}, I., {Grillo}, C., {et~al.} 2014, The Messenger, 158,
  48

\bibitem[{{Sanchis} {et~al.}(2004){Sanchis}, {Mamon}, {Salvador-Sol{\'e}}, \&
  {Solanes}}]{Sanchis+04}
{Sanchis}, T., {Mamon}, G.~A., {Salvador-Sol{\'e}}, E., \& {Solanes}, J.~M.
  2004, \aap, 418, 393

\bibitem[{{Sarazin}(1986)}]{Sarazin86}
{Sarazin}, C.~L. 1986, Reviews of Modern Physics, 58, 1

\bibitem[{{Sartoris} {et~al.}(2020){Sartoris}, {Biviano}, {Rosati}, {Mercurio},
  {Grillo}, {Ettori}, {Nonino}, {Umetsu}, {Bergamini}, {Caminha}, \&
  {Girardi}}]{Sartoris+20}
{Sartoris}, B., {Biviano}, A., {Rosati}, P., {et~al.} 2020, \aap, 637, A34

\bibitem[{{Solanes} \& {Salvador-Sol\'e}(1990)}]{SSS90}
{Solanes}, J.~M. \& {Salvador-Sol\'e}, E. 1990, \aap, 234, 93

\bibitem[{{Springel} {et~al.}(2005){Springel}, {White}, {Jenkins}, {Frenk},
  {Yoshida}, {Gao}, {Navarro}, {Thacker}, {Croton}, {Helly}, {Peacock}, {Cole},
  {Thomas}, {Couchman}, {Evrard}, {Colberg}, \& {Pearce}}]{Springel+05}
{Springel}, V., {White}, S. D.~M., {Jenkins}, A., {et~al.} 2005, \nat, 435, 629

\bibitem[{{Tiret} {et~al.}(2007){Tiret}, {Combes}, {Angus}, {Famaey}, \&
  {Zhao}}]{Tiret+07}
{Tiret}, O., {Combes}, F., {Angus}, G.~W., {Famaey}, B., \& {Zhao}, H.~S. 2007,
  \aap, 476, L1

\bibitem[{{Tonnesen}(2019)}]{Tonnesen19}
{Tonnesen}, S. 2019, \apj, 874, 161

\bibitem[{{Umetsu} {et~al.}(2014){Umetsu}, {Medezinski}, {Nonino}, {Merten},
  {Postman}, {Meneghetti}, {Donahue}, {Czakon}, {Molino}, {Seitz}, {Gruen},
  {Lemze}, {Balestra}, {Ben{\'{\i}}tez}, {Biviano}, {Broadhurst}, {Ford},
  {Grillo}, {Koekemoer}, {Melchior}, {Mercurio}, {Moustakas}, {Rosati}, \&
  {Zitrin}}]{Umetsu+14}
{Umetsu}, K., {Medezinski}, E., {Nonino}, M., {et~al.} 2014, \apj, 795, 163

\bibitem[{{Umetsu} {et~al.}(2018){Umetsu}, {Sereno}, {Tam}, {Chiu}, {Fan},
  {Ettori}, {Gruen}, {Okumura}, {Medezinski}, {Donahue}, {Meneghetti}, {Frye},
  {Koekemoer}, {Broadhurst}, {Zitrin}, {Balestra}, {Ben{\'\i}tez}, {Higuchi},
  {Melchior}, {Mercurio}, {Merten}, {Molino}, {Nonino}, {Postman}, {Rosati},
  {Sayers}, \& {Seitz}}]{Umetsu+18}
{Umetsu}, K., {Sereno}, M., {Tam}, S.-I., {et~al.} 2018, \apj, 860, 104

\bibitem[{{Valk} \& {Rembold}(2025)}]{VR25}
{Valk}, G.~A. \& {Rembold}, S.~B. 2025, \mnras, 536, 2730

\bibitem[{{Valluri} {et~al.}(2007){Valluri}, {Vass}, {Kazantzidis}, {Kravtsov},
  \& {Bohn}}]{Valluri+07}
{Valluri}, M., {Vass}, I.~M., {Kazantzidis}, S., {Kravtsov}, A.~V., \& {Bohn},
  C.~L. 2007, \apj, 658, 731

\bibitem[{{Wojtak} \& {{\L}okas}(2010)}]{WL10}
{Wojtak}, R. \& {{\L}okas}, E.~L. 2010, \mnras, 408, 2442

\bibitem[{{Wojtak} {et~al.}(2009){Wojtak}, {{\L}okas}, {Mamon}, \&
  {Gottl{\"o}ber}}]{Wojtak+09}
{Wojtak}, R., {{\L}okas}, E.~L., {Mamon}, G.~A., \& {Gottl{\"o}ber}, S. 2009,
  \mnras, 399, 812

\bibitem[{{Wojtak} {et~al.}(2008){Wojtak}, {{\L}okas}, {Mamon},
  {Gottl{\"o}ber}, {Klypin}, \& {Hoffman}}]{Wojtak+08}
{Wojtak}, R., {{\L}okas}, E.~L., {Mamon}, G.~A., {et~al.} 2008, \mnras, 388,
  815

\bibitem[{{Zarattini} {et~al.}(2021){Zarattini}, {Biviano}, {Aguerri},
  {Girardi}, \& {D'Onghia}}]{Zarattini+21}
{Zarattini}, S., {Biviano}, A., {Aguerri}, J.~A.~L., {Girardi}, M., \&
  {D'Onghia}, E. 2021, \aap, 655, A103

\end{thebibliography}

\begin{appendix}
\onecolumn
\section{Additional figures}
\begin{figure*}[ht!]
\centering
\includegraphics[width=\hsize]{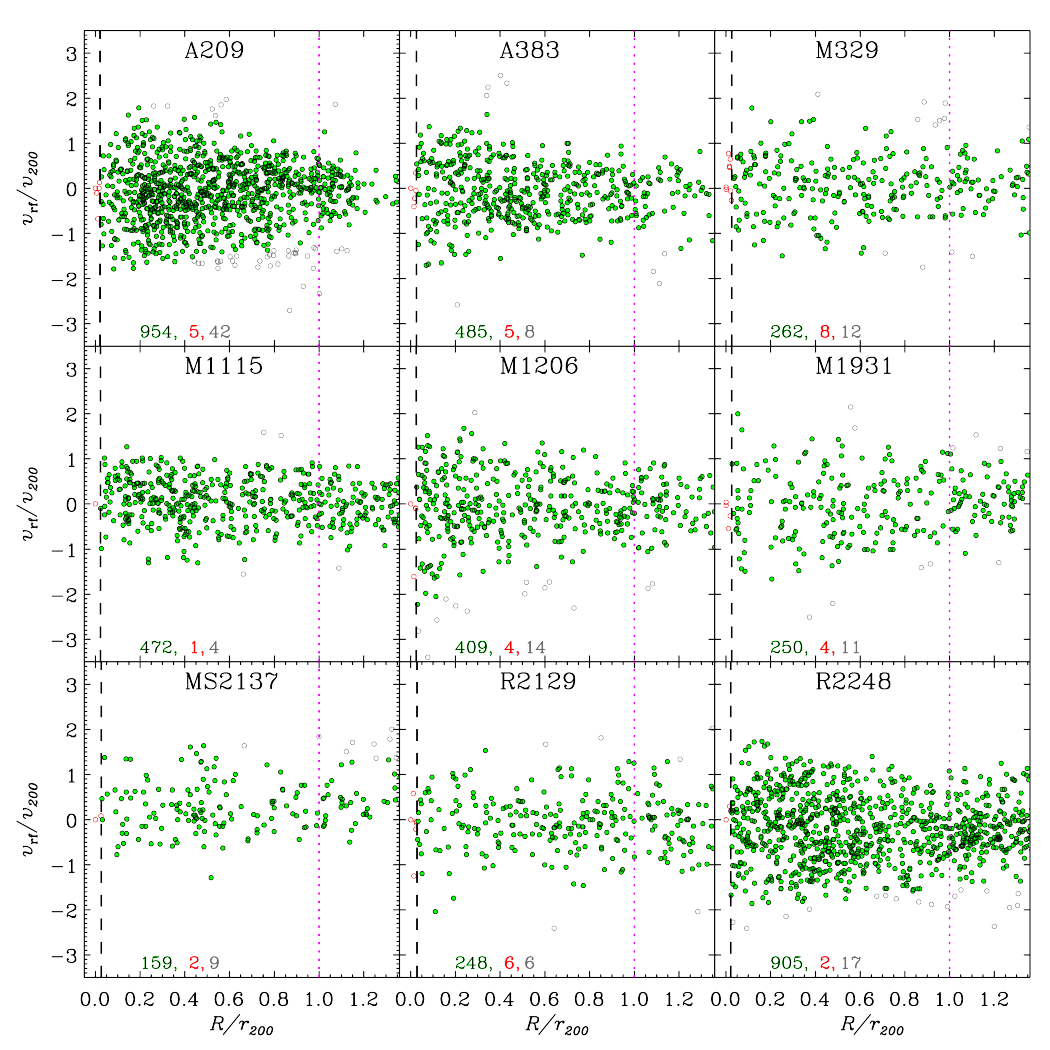}
\caption{Projected phase-space diagrams of the nine clusters. The meaning of the symbols is the same as in Fig.~\ref{f:pps1}.}
\label{f:pps}
\end{figure*}

\begin{figure}
\centering
\includegraphics[width=\hsize]{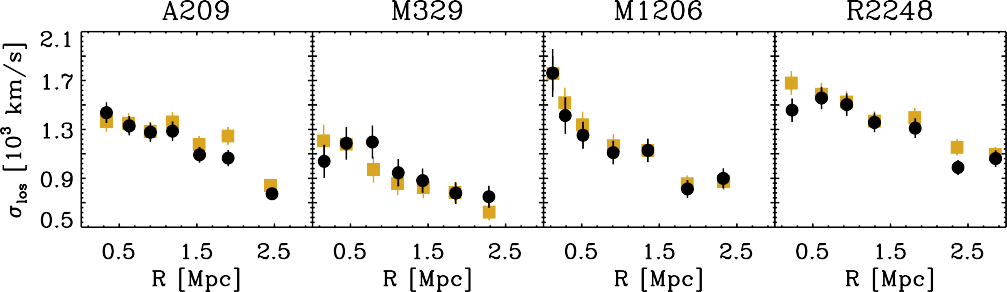}
\caption{Line-of-sight velocity dispersion profiles of four clusters in our sample, based on our membership selection on {m$_{\rm R}$} magnitude cut samples (black dots) and on previous membership selections without magnitude cuts (gold squares) by \citet{Annunziatella+16,Girardi+24,Biviano+23,Sartoris+20} for A209, M329, M1206, and R2248, respectively. The error bars are 1$\sigma$. 
}
\label{f:vdpcfr}
\end{figure}

\begin{figure*}
\centering
\includegraphics[width=\hsize]{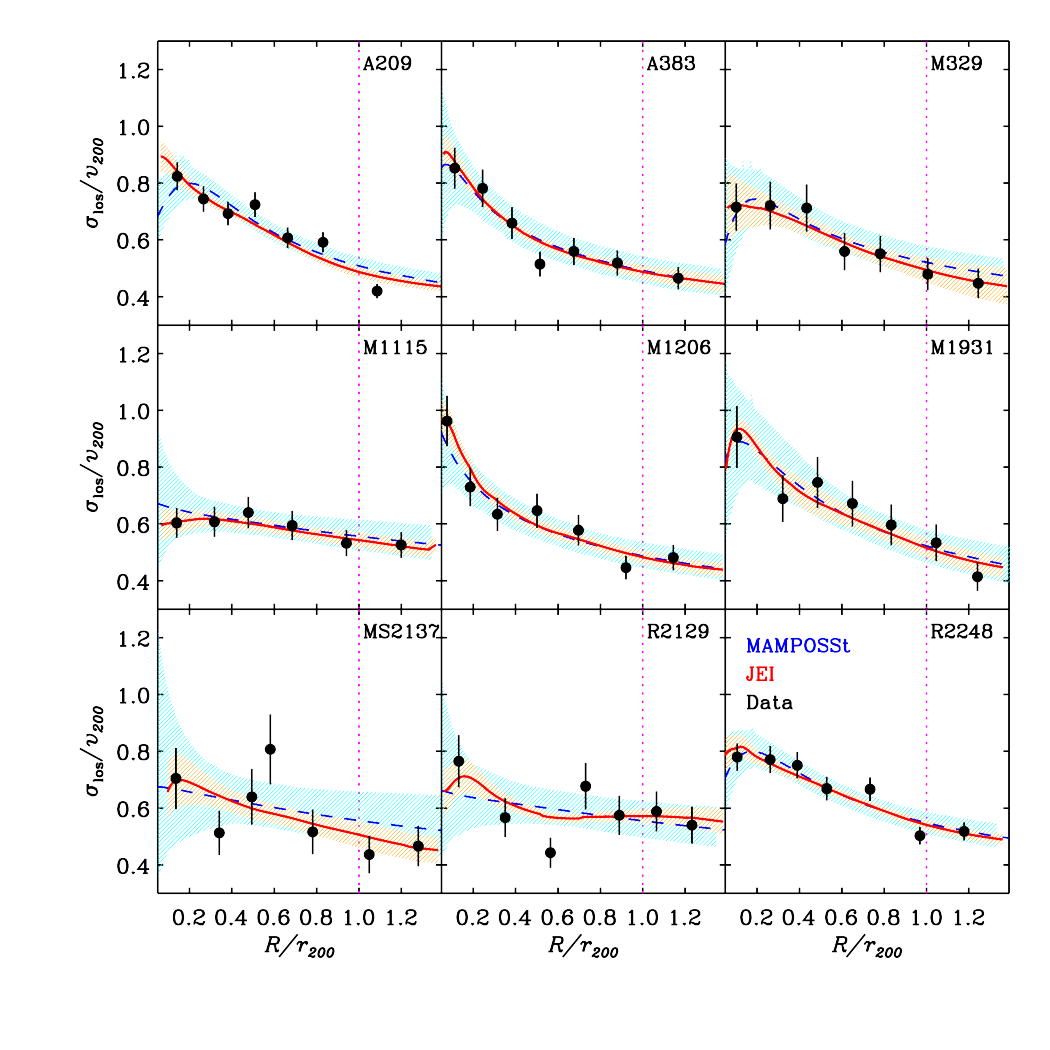}
\caption{VDPs of the nine clusters. The meaning of the symbols is the same as in Fig.~\ref{f:vdp1}. The \mamp\ solution for all clusters is that obtained for the gOM $\br$ model, except for M1931 for which the gT model is used, since it gives a better fit than the gOM model.
}
\label{f:vdp}
\end{figure*}

\begin{figure*}
\centering
\includegraphics[width=\hsize]{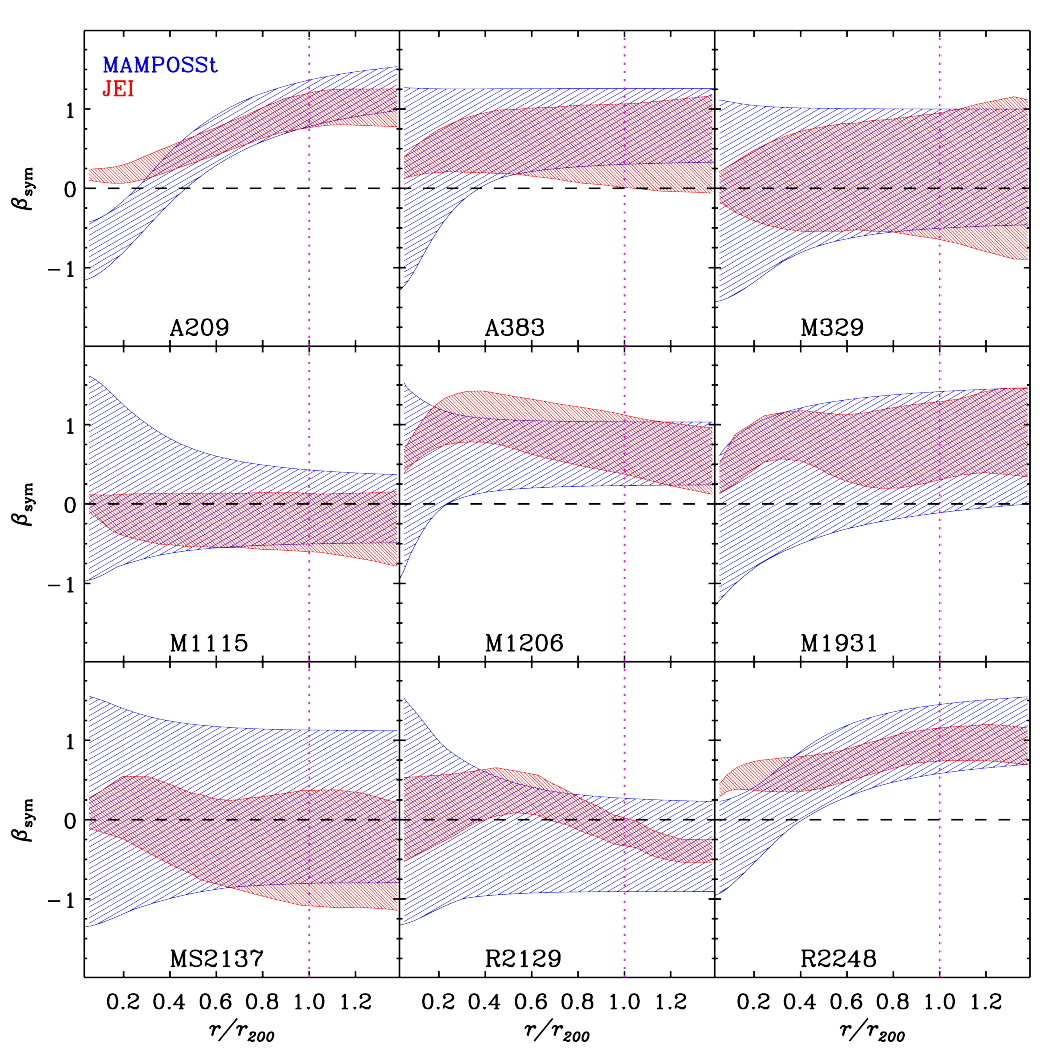}
\caption{68\% confidence regions for the \mamp\ $\bs(r)$ (blue) and for the \jei\ $\bs(r)$ (red), for the nine clusters of our sample. The \mamp\ solution for all clusters is that obtained for the gOM $\br$ model, except for M1931 for which the gT model is used, since it gives a better fit than the gOM model. The dashed horizontal line indicates isotropic orbits. Orbits are radial (respectively tangential) above (respectively below) this line.} 
\label{f:betajeimamp}
\end{figure*}

\begin{figure*}
\centering
\includegraphics[width=\hsize]{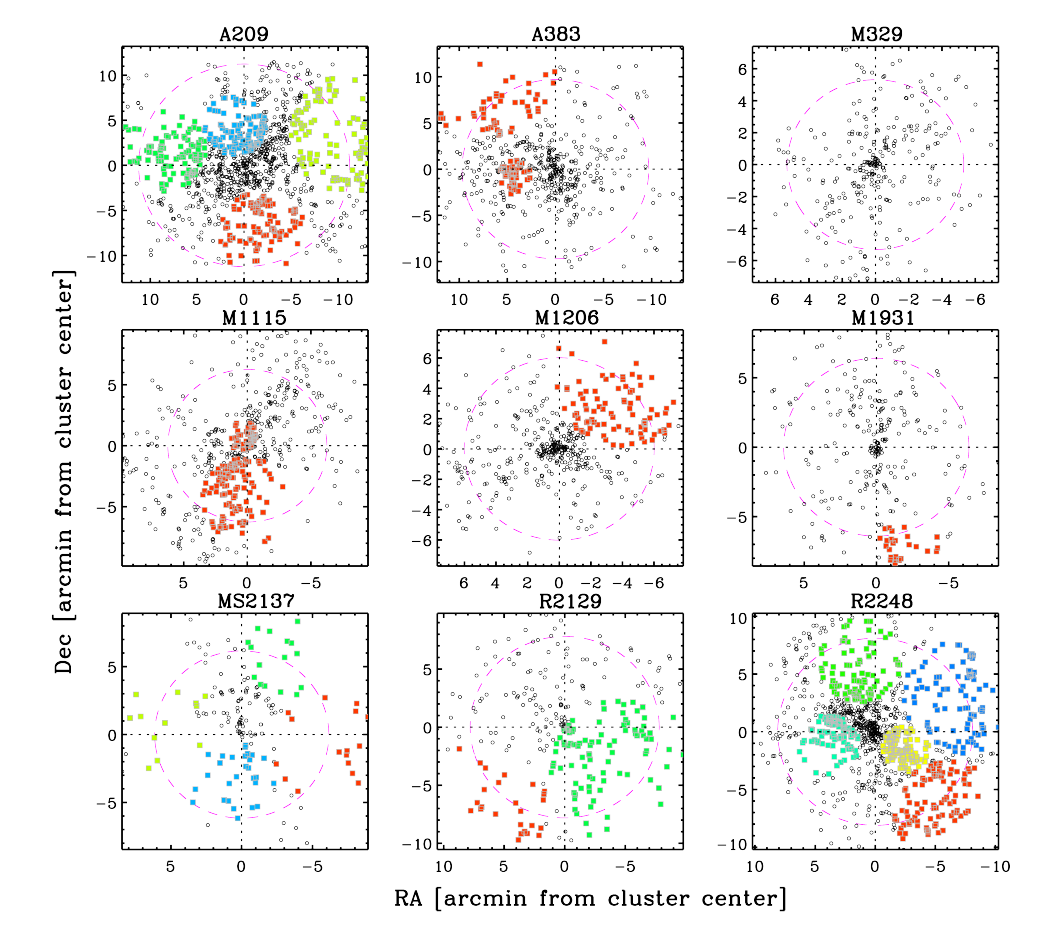}
\caption{Subclusters identified with the \dsp\ method in the nine clusters. Galaxies assigned to subclusters are shown as gray squares filled with different colors to distinguish the different subclusters. Cluster members outside subclusters are shown as black diamonds. No subcluster was identified in M329. The magenta dashed circle represents the $\rvir$ region, and the dotted lines identify the cluster center.} 
\label{f:subs}
\end{figure*}

\end{appendix}

\end{document}